\documentclass[11pt]{article}
\usepackage[affil-it]{authblk}
\usepackage{enumitem} 
\usepackage[margin=2cm]{geometry} 
\usepackage{lineno}
\usepackage{setspace}
\usepackage[utf8]{inputenc}
\usepackage[british]{babel} 
\usepackage[scaled]{helvet}
\usepackage[T1]{fontenc}
\usepackage{url}
\usepackage[colorlinks=true, allcolors=blue]{hyperref} 
\usepackage{amsmath,amsfonts,amssymb} 
\usepackage{mathpazo} 
\usepackage{siunitx}
\usepackage{graphicx}
\usepackage{caption}
\usepackage{multicol}
\usepackage{multirow}
\usepackage{threeparttable}
\usepackage[super,sort&compress,numbers]{natbib}
\usepackage{etoolbox}
\AtBeginEnvironment{thebibliography}{\linespread{1}\selectfont}

\begin{document}
\title{Quantitative characterisation of the layered structure within lithium-ion batteries using ultrasonic resonance}
\author{Ming Huang$^a$, Niall Kirkaldy$^b$, Yan Zhao$^c$, Yatish Patel$^b$, Frederic Cegla$^a$, Bo Lan$^a$\footnote{\raggedright{Corresponding author. E-mail addresses: m.huang16@imperial.ac.uk (M. Huang), n.kirkaldy@imperial.ac.uk (N. Kirkaldy), yan.zhao10@imperial.ac.uk (Y. Zhao), yatish.patel@imperial.ac.uk (Y. Patel), f.cegla@imperial.ac.uk (F. Cegla), bo.lan@imperial.ac.uk (B. Lan)}. Telephone number: +44 20 7594 7227 (B. Lan).}}
\affil{$^a$ The Non-Destructive Evaluation Group, Department of Mechanical Engineering, Imperial College London, London SW7 2AZ, UK

$^b$ The Electrochemical Science and Engineering Group, Department of Mechanical Engineering, Imperial College London, London SW7 2AZ, UK

$^c$ Breathe Battery Technologies Limited, London SE1 7SJ, UK}
\date{\vspace{-5ex}}
\maketitle

\section*{Abstract}

Lithium-ion batteries (LIBs) are becoming an important energy storage solution to achieve carbon neutrality, but it remains challenging to characterise their internal states for the assurance of performance, durability and safety. This work reports a simple but powerful non-destructive characterisation technique, based on the formation of ultrasonic resonance from the repetitive layers within LIBs. A physical model is developed from the ground up, to interpret the results from standard experimental ultrasonic measurement setups. As output, the method delivers a range of critical pieces of information about the inner structure of LIBs, such as the number of layers, the average thicknesses of electrodes, the image of internal layers, and the states of charge variations across individual layers. This enables the quantitative tracking of internal cell properties, potentially providing new means of quality control during production processes, and tracking the states of health and charge during operation.

\noindent\textbf{Keywords:} Lithium-ion battery; ultrasonic resonance; characterisation; layer structure; state of charge

\section{Introduction}

Lithium-ion batteries (LIBs) are already ubiquitous in electric vehicles, consumer electronics, and energy storage devices \cite{yoshio2009lithium}, and their usages are expected to be boosted even further by the upcoming governmental bans on fossil-fuel vehicle sales in many countries \cite{govuk_2020, eu_ice}. Manufacturers are thus incentivised to ramp up production and push performance of the batteries, for larger capacity, faster charging speeds, and lower costs \cite{nitta2015li}. However, LIB safety can still be a concern, as exposed by the high-profile Samsung cellphone failures \cite{samsungnote7, batteries4010003} and EVs catching fire \cite{tesla2019, porche_fire}.

The push for performance demands reliable characterisation and monitoring of states of charge (SOC) and health (SOH) of the batteries, while the assurance of safety requires detection and elimination of manufacturing faults (e.g. misaligned electrodes and poor cell constructions) during production \cite{robinson2020identifying}. Commonly deployed methods to infer battery states are based on electrical signals \cite{zhang2011review}, such as open-circuit voltage, internal resistance \cite{chen2018new} and capacity change \cite{li2020state}, but these indirect estimations are purely model-based and can be prone to inaccuracies.  Lab-based X-ray has been employed to ensure safety in production \cite{samsung2}, while X-ray synchrotron \cite{mcbreen2009application, finegan2015operando, vanpeene2019dynamics} and computational tomography \cite{ebner2013visualization, frisco2016understanding, taiwo2017investigating} have become powerful tools in research and development, but their radiation hazards, high costs and practical limitations have also limited their wider deployment.

Ultrasound has been recognised as an attractive candidate for rapid, cheap and accessible examination of LIBs. The active materials within LIB electrodes undergo changes in physical properties (e.g. wave speed and density) during normal operation. The structure of the electrodes can also change as cells age due to the effects of various degradation mechanisms. These changes can be detected by ultrasound non-destructively. Indeed, this has spurred considerable research employing various acoustic techniques. For instance, Villevieille et al. \cite{villevieille2010direct} used acoustic emission to investigate in-operando the structural and morphological changes of the electrodes. Ladpli et al. \cite{ladpli2018estimating} demonstrated the feasibility of monitoring the SOC and SOH using acoustic guided waves. Recently, the conventional time-of-flight (TOF) measurements using the ultrasonic through-transmission or pulse-echo configurations have been adopted in a range of investigations. The pioneer work by Hsieh et al. \cite{hsieh2015electrochemical} showed clear correlations using experimental TOF measurements of 2.25MHz ultrasound across all SOCs. Gold et al. \cite{gold2017probing} investigated the same issue with a much lower frequency (200kHz), and proposed the second Biot mode in porous materials as the theoretical model to explain the slow wave propagation speed. In contrast, Davies et al. \cite{davies2017state} used the classic Hashin-Shtrikman homogenisation method for composites \cite{HASHIN1963127} to model the wave propagation through electrodes, and applied machine learning to successfully extract SOC and SOH from ultrasonic data. Similar experimental and theoretical frameworks were then successfully applied to detect lithium metal plating \cite{bommier2020operando} and to measure the effective stiffness of the battery \cite{chang2020measuring}, among other things. Meanwhile, Robinson et al. \cite{robinson2019spatially} highlighted the need and advantages for such TOF measurements of battery states and stiffnesses to be carried out in a spatially resolved fashion.

These studies have comprehensively proved the correlation between ultrasound and the battery internal states. However, the fact that they are mostly based on the through-transmission measurements of ultrasound means that they are only able to deliver average estimations across the thickness without spatial resolution. Given that a battery is a composite system of tens of layers with contrasting properties, such averages of properties can be rather crude and have limited accuracy. In addition, their common physical assumption that electrodes are porous materials have actually found limited success against experimental results (more discussions in section \ref{sec:layers}).

Recently there have been efforts to achieve spatial resolution in the thickness direction. For example, Robinson et al. \cite{robinson2020identifying} demonstrated that a certain reflection peak between the main front- and back-wall echoes, or the lack of it, can be associated to a missing layer of an artificially constructed battery, and used to image the profile of the said layer. Similarly, by applying acoustic microscopy on a battery and gating the time-domain signal to a certain range, Bauermann et al. \cite{BAUERMANN} were able to obtain high-quality images of the inner structures such as a fine mesh. Even though these works mainly focused on the detection of features/defects instead of characterisation of properties, they did point to the extra information carried by the time-domain waveforms along the thickness.

We present a new methodology for layer-resolved characterisation of LIBs using ultrasonic resonance, and its novelties are threefold. Firstly, it employs simple, conventional ultrasound equipment to robustly acquire the resonant time traces, which, compared to the transmitted signal mostly used so far in literature, inherently carry more information. Secondly, a pivotal theoretical model is established, from ground up, to comprehensively illustrate the fundamental wave physics on how the resonance is formed from the reflections from the repetitive internal layers. This key contribution opens the door for new quantitative characterisation capabilities of the battery layers using ultrasound, which are demonstrated in a range of applications as the last aspect of the novelties. 

The paper is organised to clearly explain the methodology and highlight the contributions: section \ref{sec:experiments} firstly demonstrates robust, general experimental observations of resonance. Then section \ref{sec:methods} introduces the theoretical model to explain the wave physics within each layer and the formation of ultrasonic resonance from inter-layer reflections. The exciting possibilities of quantitative characterisations of internal cell structures and states are then illustrated by a variety of case studies in section \ref{sec:results}. These include estimating the number of layers, determining anode and cathode thicknesses, constructing the image of individual layers in the thickness direction, and tracking layer-level SOC changes during charging cycles. Section \ref{sec:conclusions} concludes this paper.

\section{Experimental observation of ultrasonic resonance}\label{sec:experiments}
 
\begin{figure}[!ht]
    \centering
    \includegraphics[width=\linewidth]{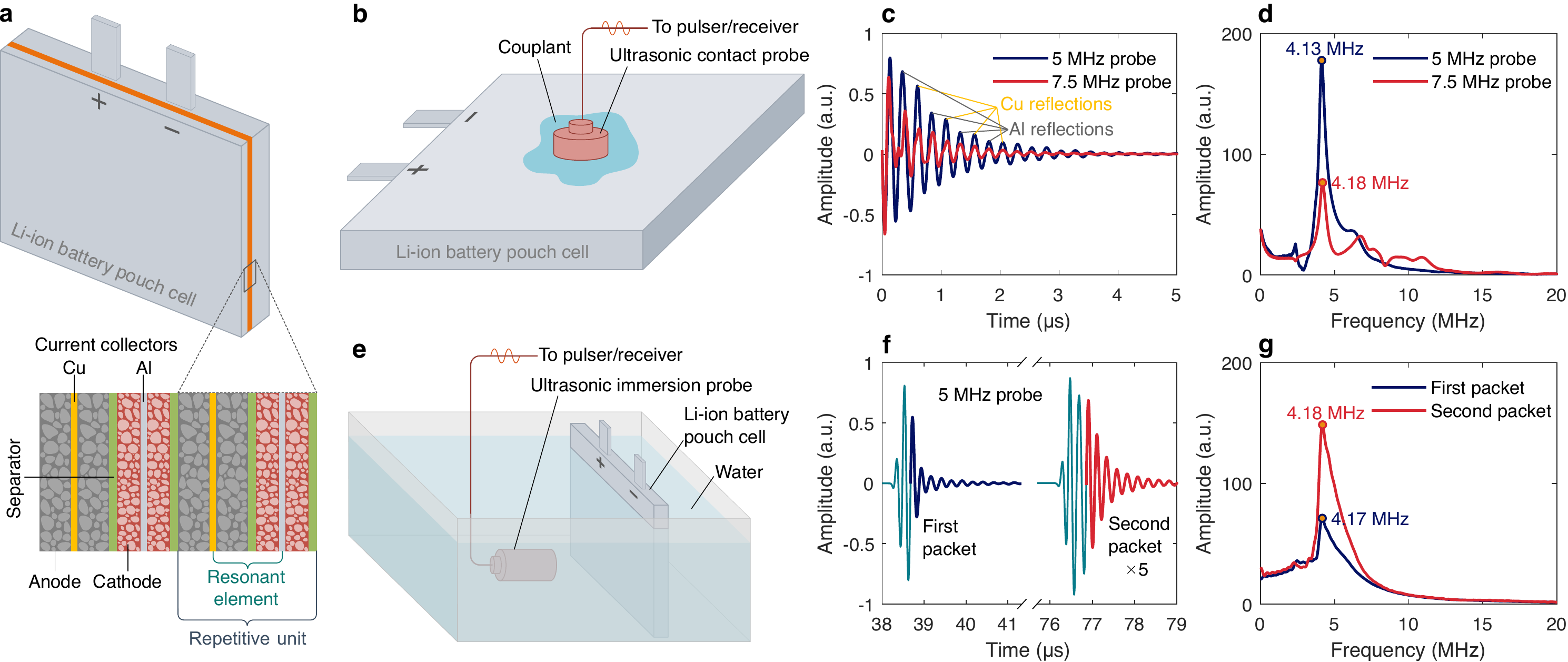}
    \caption{Observation of ultrasonic resonance from contact and immersion tests of an LIB pouch cell. \textbf{a} illustrates the pouch cell and its two repetitive internal units. The fundamental element for forming ultrasonic resonance is also highlighted. \textbf{b} illustrates the contact setup with a probe controlled by a pulser/receiver, to generate waves into the cell and receive the reflections. Water-based gel is used as couplant to facilitate wave transmission. \textbf{c} shows the time-domain signals of 5 MHz and 7.5 MHz contact probes from the Kokam 7.5 Ah cell (example resonance peaks, with amplitudes alternately slightly higher or lower than an exponential decay, are respectively from Cu and Al layers, with explanations later), while \textbf{d} displays the respective amplitude spectra. \textbf{e} shows the immersion setup with the cell partly immersed in water. \textbf{f} presents the signal from the same cell acquired by a 5 MHz immersion probe placed 28.5 mm away from the cell, with the two wave packets representing the first and second reflections. Light blue cycles are dominated by the packaging layer, while dark blue and red ones indicate the main resonance. \textbf{g} shows the amplitude spectra of the two packets after removing the cycles (light blue parts in \textbf{f}) that are dominated by front-wall reflections.}
    \label{fig:experiments}
\end{figure}

The main experimental sample, a Kokam 7.5 Ah pouch cell (SLPB75106100), is a typical LIB cell as illustrated in Fig. \ref{fig:experiments}a. It has a periodic repetition of internal layers, with each repetitive unit consisting of one Cu and one Al current collector, two anodes and two cathodes, and one separator. In this paper, the anode and cathode electrodes refer to the layer of active material coated on the metal current collector. The thickness of each internal layer has been destructively measured in the literature \cite{Ecker2015, Hales2019}, as summarised in Table \ref{tab:properties}.

We performed ultrasonic tests on the cell in a pulse-echo configuration using the contact and immersion setups in Fig. \ref{fig:experiments}b and e. The contact setup had a 5 MHz (V109-RM, Olympus) and a 7.5 MHz (V121-RM, Olympus) probe in direct contact with the cell, with water-based gel used as couplant to facilitate wave transmission. The tests delivered the time-domain signals in Fig. \ref{fig:experiments}c, with their respective frequency-domain amplitude spectra (calculated via Fourier transform) in Fig. \ref{fig:experiments}d. The immersion setup used a 5 MHz immersion probe (Harisonic I3-0504-S; default immersion probe throughout this work) to examine the cell partly immersed in water. It received two reflected wave packets shown in Fig. \ref{fig:experiments}f, which correspond to the first and second reflections from the cell, with their respective amplitude spectra in Fig. \ref{fig:experiments}g. Note that to process the time traces in Fig. \ref{fig:experiments}f for amplitude spectra, the first several cycles (light blue parts) need to be cut out, since they are generally dominated by the front-wall (i.e. the outer surface of the cell facing the probe, whereas the back-wall is the opposite surface of the cell) instead of internal reflections. 

Comparing the independent experimental results in Fig. \ref{fig:experiments}d and g, it becomes clear that despite their different experimental setups, probes, and centre frequencies, all measurements gave very similar spectra with a pronounced resonance of practically the same frequency. These confirm that the resonance was indeed from the internal structures within cell.

We subsequently investigated the generality of the resonance, by carrying out the same experiments on other LIB pouch cells, including a 210 mAh cell (PL-651628-2C, AA Portable Power Corp.) that had been previously evaluated \cite{davies2017state,Robinson2019,Pham2020} and a Kokam 5 Ah cell (SLPB11543140H5) that had been destructively analysed \cite{Hunt2016,Hales2019}. The acquired time- and frequency-domain results are shown in Fig. \ref{fig:more_experiments}, which displayed strong resonance similar to those in Fig. \ref{fig:experiments}. These confirm that the ultrasonic resonance is not a property of one cell only, but instead can be obtained in general cases.

\section{Wave physics of the ultrasonic resonance}\label{sec:methods}

The experimentally observed resonance originates from reflections from the repetitive layers within the battery cell. To explain the fundamental mechanisms of this formation, three key components are outlined in this section: the wave propagation in the separator and electrode layers, the reverberations and total wave responses from a metal current collector layer, and the formation of resonance by the interference of reflections. These are covered consecutively in the three subsections below.

\subsection{Wave propagation in separator and electrode layers}\label{sec:layers}

In the experiments, ultrasonic waves propagate within an individual layer at a characteristic wave speed $c$, which is determined by the elastic properties and density $\rho$ of the layer material. Across the cell, the wave goes through layers with different properties, including metal sheets (including Cu and Al current collectors and packaging layers), separators and electrodes; their free-propagation speeds are provided in Table \ref{tab:properties} (though the thin metal sheets cause wave reverberations - details in the next subsection). The wave speed in separator is calculated via the classic Biot model for fluid-saturated porous media \cite{biot1,biot2}, which accounts for the fact that separators are usually porous polymers filled with liquid electrolytes \cite{Martinez-Cisneros2016}, acting as an ionically-conductive physical barrier between two electrodes. The calculation details are provided in Appendix \ref{appdx:separator}.

\begin{table}[ht]
    \centering
    \begin{threeparttable}
    \caption{Layer properties of the Kokam 7.5 Ah cell. Units: speed $c$ (m/s), density $\rho\,\mathrm{(kg/m^3)}$, impedance $Z\,\mathrm{(Pa\cdot s/m^3)}$, thickness $d\,(\SI{}{\micro\meter})$, number of layers $n$ (1).}
    \label{tab:properties}
    \begin{tabular}{llllll}
        \hline
        Layer & $c$ & $\rho$ & $Z$ & $d$ & $n$ \cite{Ecker2015}\\
        \hline
        Cu & 4762 \cite{Ledbetter1980} & 8940 \cite{Ledbetter1980} & 4.26$\times10^7$ & 14.7 \cite{Ecker2015} & 24\\
        Al & 6346$^{*}$ & 2700 \cite{DaviesThesis} & 1.71$\times10^7$ & 15.1 \cite{Ecker2015} & 25\\
        Separator & 1209 & 1063 & 1.29$\times10^6$ & 19.0 \cite{Ecker2015} & 50\\
        Anode (SOC=0)$^\ddag$ & 1341 & 1909 & 2.56$\times10^6$ & 64.2$^\dag$ (73.7 \cite{Ecker2015}) & 48\\
        Cathode (SOC=0)$^\ddag$ & 1093 & 4172 & 4.56$\times10^6$ & 47.5$^\dag$ (54.5 \cite{Ecker2015}) & 48 \\
        Anode (SOC=1)$^\ddag$ & 1443 & 1994 & 2.88$\times10^6$ & 67.4$^{\dag\dag}$ (73.7 \cite{Ecker2015}) & 48 \\
        Cathode (SOC=1)$^\ddag$ & 1136 & 3848 & 4.37$\times10^6$ & 47.5$^\dag$ (54.5 \cite{Ecker2015}) & 48 \\
        Casing (Al) & 6346 & 2700 & 1.71$\times10^7$ & 110$^\S$ & 2\\
        Whole cell & \multicolumn{4}{l}{1360 (predicted from layers. Experimentally measured: 1404)}\\
        \hline
    \end{tabular}
    \begin{tablenotes}
        \item[$^{*}$] Calculated with the self-consistent theory \cite{Kube2016} using elastic constants from Qi et al. \cite{Qi2014}
        \item[$^\dag$] If calculated directly from the destructively obtained electrode layer thicknesses, the total thickness of the cell (8.05 mm) would exceed the actual value of 7.26 mm. Thus, we have proportionally scaled those electrode values (metal layers and separator  unchanged) for the total thicknesses to match. \item[$^\ddag$]The scaling of thicknesses also means the electrode porosity values need to be adjusted, since the total volumes of the solid phases should remain consistent. The values used here are 0.23 for anode and 0.19 for cathode, which are derived from experimental results \cite{Ecker2015} with details below.
        \item[$^{\dag\dag}$] This thickness has also considered a 5\% expansion at SOC=1 compared to SOC=0 (the expansion of cathode layer is smaller than 1\% and is thus neglected here) \cite{Koyama2006}. 
        \item[$^\S$] Measured in this work.
    \end{tablenotes}
    \end{threeparttable}
\end{table}

One potential contribution of this paper is the theoretical modelling of electrodes. They are often treated as porous solids in literature, but the wave speeds predicted by the Biot \cite{biot1,biot2} or composite homogenisation \cite{HASHIN1963127} models (> 3000 m/s \cite{gold2017probing, davies2017state}) are in poor agreement with typical experimental results (<1800 m/s \cite{chang2020measuring}). To explain this, Gold et al. \cite{gold2017probing} proposed that the experiments actually obtained the shear mode wave of the Biot model, but it does not explain why the first compressional mode, which should be much easier to detect \cite{plona1980observation}, were not observed in our experiments. These suggest that the porous solid assumption may not be fundamentally accurate. Here we demonstrate that a slurry model can deliver much closer estimations of the electrode material properties than the homogenisation models. Physically, this is based on the observations that the experimental compressional wave speeds are closer to that of the liquid electrolyte than to those of the solid active materials (unlike porous solid model predictions), and that the solid particles in electrodes are spatially separate and are only held together by the liquid-saturated soft polymer binder lattice \cite{lu20203d}, whose binding force is likely the Van Der Waals force \cite{ma2019microrheological}. From mechanics point of view, the latter observation is different from typical rigid porous solids, e.g. ceramic or cured cement, whose solid phases are atomically bound together and are thus much stronger; rather, electrodes saturated in the liquid electrolyte may have more similarities with sedimented sand in water, which is a typical concentrated slurry. 

This assumption enables the estimations of the wave speeds and densities for the electrodes of the Kokam cell via a well-established theoretical model \cite{ATKINSON1992577} as detailed in Appendix \ref{appdx:slurry}. Using this model and the properties of the electrolyte and the active materials in Table \ref{tab:electrodes}, we obtained the results as listed in Table \ref{tab:properties}. Note that these estimations are sensitive to the electrode porosity (i.e. non-solid phase volume fraction), which could in turn be well-informed from simple wave speed measurements (e.g. during manufacturing). The predicted wave speeds in individual electrode layers are now shown to be dominated by the liquid phase, and the overall wave speed through our sample cell (1360 m/s) is very close to the experimental result (1404 m/s); in addition, the predicted overall speeds of a combined anode and cathode layer differ $\sim10\%$ between the fully discharged (SOC=0) and charged (SOC=1) states, which again agrees well with experiments \cite{davies2017state}. These substantiates the suitability of the slurry model for the estimation of electrode properties.

\subsection{Reverberations and total wave responses in a metal layer}

We now proceed to examine the wave interactions with the metal current collector layers. Generally, when incident upon an interface, part of the wave is reflected back in the incident medium (denoted 0) while the rest is transmitted through to the other medium (denoted 1). The amplitudes of these two wave parts are described by the well-known reflection $R_{01}$ and transmission $T_{01}$ coefficients \cite{auld1973acoustic}:
\begin{linenomath*}
\begin{equation}\label{eq:rt}
    R_{01}=\frac{Z_0-Z_1}{Z_0+Z_1},\;T_{01}=\frac{2Z_0}{Z_0+Z_1},
\end{equation}
\end{linenomath*}
where the acoustic impedance $Z = \rho c$ is a function of the wave speed $c$ and density $\rho$ of the material. Notice that the reflection and transmission depend not only on the impedance contrast across the boundary, but also on which material the wave is incident from.

\begin{figure}[ht]
	\centering
	\includegraphics[width=0.8\linewidth]{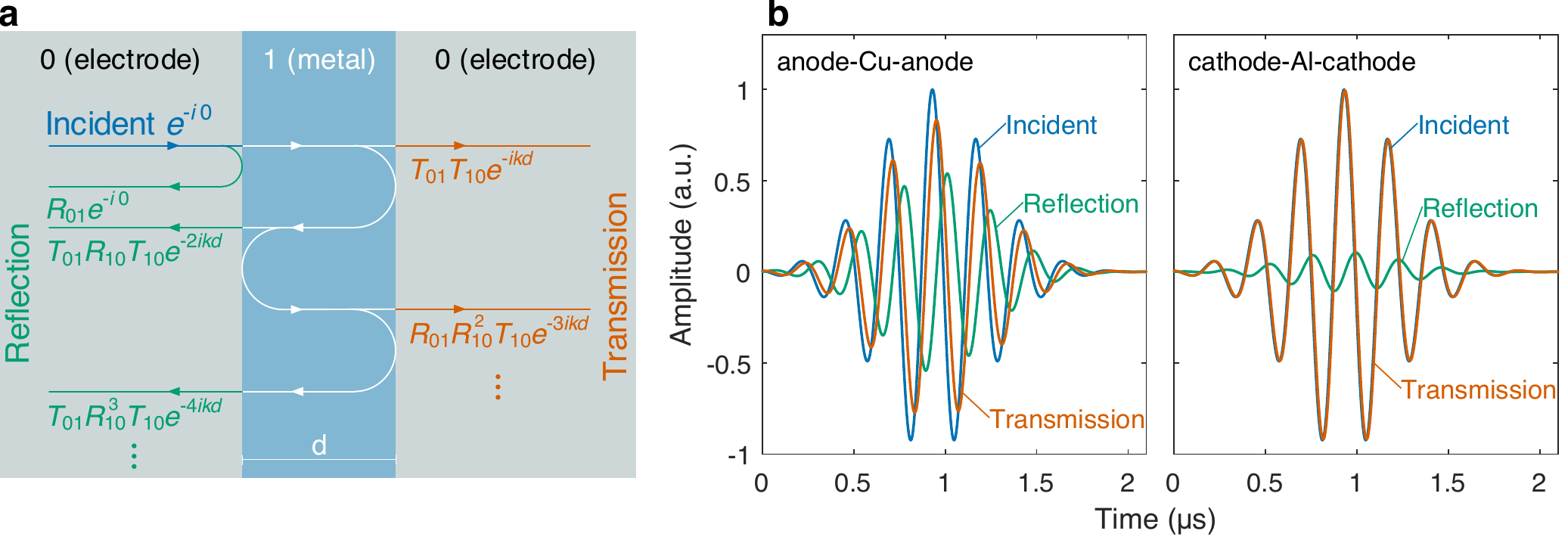}
	\caption{Ultrasonic reverberations and the resultant total signals of the metal current collectors. \textbf{a} shows the individual reflected (R) and transmitted (T) waves caused by multiple reverberations within the metal layer. Their amplitudes are respectively described by $R_{01}$ and $T_{01}$ (subscript 0 before 1) for a wave travelling from 0 to 1; each subsequent reflection is shifted by a phase related to the wave number $k$ and layer thickness $d$. \textbf{b} illustrates the total reflection and transmission (summations of an infinite number of individual reflected and transmitted waves) from the Cu and Al layers of the Kokam cell.}
	\label{fig:fabry_perot}
\end{figure}

For a thin current collector, due to the close proximity of its boundaries, the reflection and transmission at both boundaries would interfere and cause reverberations in between. As exemplified in Fig. \ref{fig:fabry_perot}a for a metal layer bonded with electrode on both sides, each reflection (or transmission) is separated from its subsequent one by a phase shift corresponding to the round-trip propagation through the layer thickness. The total macro responses of reflection $R$ and transmission $T$ from the metal layer are summations of all individual reflections and transmissions, given by\cite{Brekhovskikh1980}:
\begin{linenomath*}
\begin{equation}\label{eq:fabry_perot}
    \begin{aligned}
        R&=R_{01}+T_{01}R_{10}T_{10}e^{-2ikd}+T_{01}R_{10}^3T_{10}e^{-4ikd}+\ldots=R_{01}\frac{1-e^{-2ikd}}{1-R_{01}^2e^{-2ikd}},\\
        T&=T_{01}T_{10}e^{-ikd}+T_{01}R_{10}^2T_{10}e^{-3ikd}+\ldots=\frac{(1-R_{01}^2)e^{-ikd}}{1-R_{01}^2e^{-2ikd}},
    \end{aligned}
\end{equation}
\end{linenomath*}
with $k$ and $d$ denoting the wave number and thickness of the metal layer. 

Eq. \ref{eq:fabry_perot} shows that a strong total reflection needs both strong individual reflections (requiring large contrast of acoustic properties across the boundary) and small enough phase shifts to form approximately constructive interference (requiring thin thicknesses). Both are the case for the metal layers inside LIBs in the considered MHz frequency range, but the reverberations in electrodes and separators are negligible, due to the similar acoustic impedances between the neighbouring anode-separator-cathode layers, as can be seen from Table \ref{tab:properties}. Therefore, they can be effectively treated as a combined thicker layer, within which a sound wave travels freely.

As a result, we only consider the reverberations in metal layers and neglect those in electrodes and separators. For the Cu and Al layers of our cell, the total reflected and transmitted waves calculated from Eq. \ref{eq:fabry_perot} are exemplified in Fig. \ref{fig:fabry_perot}b. They were calculated in the frequency domain using the properties in Table \ref{tab:properties}, by applying the appropriate amplitude and phase modulations on each individual frequency component of the incident wave, and transforming the results back to the time domain. It is important to point out that they both have phase shifts and amplitude changes relative to the incident wave, which are caused by the reverberations. Regarding the phase shifts, it can be mathematically proven from Eq. \ref{eq:fabry_perot} that the reflected and transmitted waves have a consistent $\pi/2$ phase difference:
\begin{linenomath*}
\begin{equation}\label{eq:rt_relation}
    \phi_T=\phi_R+\pi/2,
\end{equation}
\end{linenomath*}
which is true irrespective of the material pairs, frequency, or layer thickness. This relationship is an important step to arriving at the resonance condition in the next subsection. For the amplitude changes, a prominent observation is that the reflection from the Cu layer is much stronger in amplitude than that from the Al layer in this case. This amplitude difference is also evident from the experimental results in Fig. \ref{fig:experiments}c, which shows slightly alternating amplitudes of the peaks and could be used to distinguish the metal layers. However, we emphasise that it is entirely possible for another battery configuration to have stronger reflections from the Al layers, because it depends on various factors including the acoustic impedance mismatch between the metals and electrodes, the wave frequency and layer thickness.

\subsection{Ultrasonic resonance formation}

\begin{figure}[ht]
	\centering
	\includegraphics[width=\linewidth]{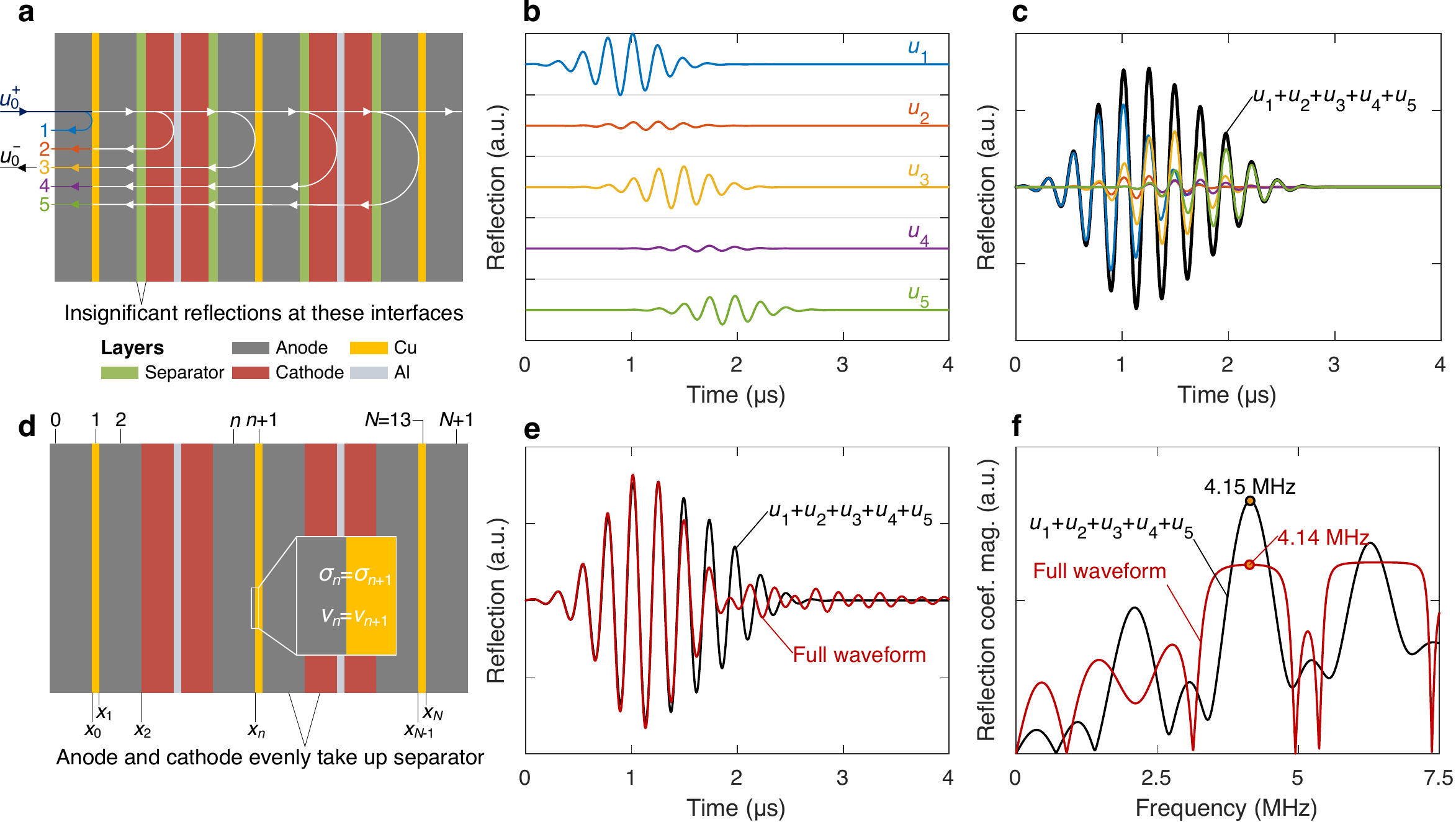}
	\caption{Formation of resonance in an LIB. \textbf{a} illustrates the major reflections that contribute to the formation of resonance. \textbf{b} displays the actual major reflections when subjecting to the incident wave shown in Fig. \ref{fig:fabry_perot}b. \textbf{c} superposes the major reflections to form the total reflection as shown in black. \textbf{d} illustrates the setup for full-waveform modelling with the multilayered battery consisting of $N$ layers, bounded by semi-infinite media 0 and $N+1$; the model is formulated by satisfying the continuity of stress, $\sigma_n=\sigma_{n+1}$, and velocity, $v_n=v_{n+1}$, across each individual interface $n$. \textbf{e} compares the full-waveform result with that predicted in \textbf{c}. \textbf{f} compares the respective reflection coefficients obtained by the two analytical models in the frequency domain.}
	\label{fig:resonance}
\end{figure}

With the wave physics in individual layers and interfaces understood, we can now derive the conditions needed for the ultrasonic resonance to form, which can be simply based on the constructive interference of major reflections from metal layers. For example, the first few major reflections and their wave paths for the Kokam cell are illustrated in Fig. \ref{fig:resonance}a. As discussed, the reflected and transmitted signals from a metal layer have implicitly accounted for the infinite reverberations within it, while only a single-trip transmission is considered for each anode-separator-cathode combined layer.

Due to their different wave paths, the major reflections incur different phase shifts compared to the incident wave. For a resonance to form at a given frequency, these reflections need to have an exact $2n\pi$ phase difference between them, with the integer number $n$ denoting the order of resonance. For instance, this requires the phase shifts $\phi_{u_1}$ and $\phi_{u_2}$ of the first two reflections to satisfy:
\begin{linenomath*}
\begin{equation}\label{eq:resonance0}
    \phi_{u_1}=\phi_{u_2}+2n\pi
\end{equation}
\end{linenomath*}
The two phase shifts are given by $\phi_{u_1} = \phi_\mathrm{rc}$ and $\phi_{u_2} = 2\phi_\mathrm{tc}+2\phi_\mathrm{e}+\phi_\mathrm{ra}$, where $\phi_\mathrm{e}$ is the phase shift in the combined anode-separator-cathode layer, and $\phi_\mathrm{rc}$, as an example, is the reflection (first subscript, $r$ for reflection and $t$ for transmission) from the Cu (second subscript, $c$ for Cu and $a$ for Al) layer. Substituting these expressions and, importantly, Eq. \ref{eq:rt_relation}, into Eq. \ref{eq:resonance0} results in the defining relation for ultrasonic resonance:
\begin{linenomath*}
\begin{equation}\label{eq:resonance}
    \phi_\mathrm{ra}+ 2\phi_\mathrm{e}+\phi_\mathrm{rc} = -(2n+1)\pi.
\end{equation}
\end{linenomath*}

Eq. \ref{eq:resonance} is formulated based on $u_1$ and $u_2$, but it can be proven that the same equation would be similarly obtained from the phase relationship between $u_2$ and $u_3$. These mean that $u_1$ and $u_3$ would automatically satisfy the $2n\pi$ phase shift and form resonance, and so will all subsequent reflections that have gone through a single round trip (i.e. transmitted through multiple layers and only reflected once), e.g. between $u_2$ and $u_4$. It thus transpires that the simple equation, Eq. \ref{eq:resonance}, in fact governs the general condition of resonance for any reflection pair, regardless of whether the equations are constructed with Cu as the first layer (as above) or Al. 

Moreover, the resonance described in Eq. \ref{eq:resonance} originates from a Cu layer and its neighbouring Al layer (e.g. $u_1$ and $u_2$), which is the main focus of this paper. However, resonance can also form between a Cu layer and the next Cu layer, e.g. $u_1$ and $u_3$ and so on. These considerations can generalise Eq. \ref{eq:resonance} further into:
\begin{linenomath*}
\begin{equation}\label{eq:gene_resonance}
   \phi_\mathrm{ra}+ 2\phi_\mathrm{e}+\phi_\mathrm{rc} = -n\pi. 
\end{equation}
\end{linenomath*}
When $n$ is odd, the resonance is from Cu-Al layers; whereas when $n$ is even, the resonance is from two Cu or two Al layers. For example, the first and third peaks of the black curve in Fig. \ref{fig:resonance}f likely come from $n=0$ and $2$, respectively.  

So far we have considered all signals that make a straight single round-trip to one metal layer, and now we prove that the waves which have been reflected multiple times by the metal layers cannot form resonance. Let us consider an exemplary signal $u_3^*$ that has a similar travelling distance with the major round-trip reflection $u_3$ (interacted with metal layers 3 times), but bounces back and forth between the Al and Cu layers. For these two signals to form resonance, they need to satisfy
\begin{linenomath*}
\begin{equation} \label{eq:reverb_1}
\begin{aligned}
    2\phi_\mathrm{tc} + 4\phi_\mathrm{e} + 2\phi_\mathrm{ra}+\phi_\mathrm{rc} + 2n\pi&= 2\phi_\mathrm{tc} + 4\phi_\mathrm{e} + 2\phi_\mathrm{ta}+\phi_\mathrm{rc} \\
   \text{i.e. } \phi_\mathrm{ra} + n\pi &= \phi_\mathrm{ta}  
\end{aligned}
\end{equation}
\end{linenomath*}
Similarly, the requirement for the fourth reflections is:
\begin{linenomath*}
\begin{equation}\label{eq:reverb_2}
   \phi_\mathrm{rc} + n\pi = \phi_\mathrm{tc} 
\end{equation}
\end{linenomath*}
However, Eqs. \ref{eq:reverb_1} and \ref{eq:reverb_2} contradicts with the inherent relationship for any metal layer shown in Eq. \ref{eq:rt_relation}, which means they can never be satisfied, and that no resonance can be formed between $u_3$ and $u_3^*$. Instead, the weaker signal $u_3^*$ always has a phase difference of $\pi$ from the main reflection $u_3$, i.e. it is always destructively interfering the main, round-trip reflections and reducing the latter's amplitudes. This also means that in analysing the experimental resonance results from purely the phase relationship point of view (amplitude of $u_3^*$ is < 2$\%$ of $u_3$), as shown in Eq. \ref{eq:resonance}, the multiply reflected signals like $u_3^*$ have negligible contributions, and only the main round-trip reflections need to be considered.


Using the layer properties in Table \ref{tab:properties}, the first-order main resonance ($n=1$) of the Kokam cell at SOC=0 was estimated from Eq. \ref{eq:resonance} to be 4.15 MHz, agreeing well with experimental results. The major time-domain reflections from the first few metal layers, when subject to a 4.15 MHz centre-frequency incident signal, are illustrated individually in Fig. \ref{fig:resonance}b. They are plotted together in coloured lines of Fig \ref{fig:resonance}c, and through constructive interference, their summation (the black line) evidently forms the main resonance as received in experiments.

This simple physical model was rigorously evaluated against a full-waveform model, which takes into account all possible wave events in a cell by considering the continuity of velocity (compatibility, $v_n=v_{n+1}$) and stress (equilibrium, $\sigma_n=\sigma_{n+1}$) across each individual interface $n$ as outlined in Fig. \ref{fig:resonance}d, with details in Appendix \ref{appdx:full_waveform}. Here the reflections between separators and electrodes are not considered, since they were not observed experimentally. The time- and frequency-domain results of the two methods are compared in Fig. \ref{fig:resonance}e and f. The results demonstrate reasonably good agreements between the two methods, especially for the frequency domain. These prove that, despite its simplicity, the method in Eq. \ref{eq:resonance} can deliver very accurate predictions for the main resonant frequency. An advantage for the full-waveform model, however, is its handling of amplitude, which the simple model neglects. This has relevance for resonance amplitudes and wave attenuation, and is potentially useful in the study of electrode material degradation.

\section{Application case studies}\label{sec:results}

We have so far completed the outline of the physical model for analysing ultrasonic resonance, which opens up various characterisation opportunities. Firstly, the resonant frequency corresponds to the overall ultrasound behaviour in the cell, so it enables quantitative evaluation of average cell properties and states. Secondly, the time trace carries spatially-resolved information about individual layers in the depth direction, i.e., signal peak TOFs are associated with individual metal layers, and the reflections from Cu and Al layers are distinguishable from their amplitudes. This allows for characterisations of battery structures and states to a layer-resolved level. The two aspects of the resonance is visible in e.g. Fig. \ref{fig:experiments}d and g, where the average cell property determines the central resonant frequency, and the spatial variations contribute to the width of the frequency spectra. The exciting characterisation possibilities are explored in the following section as application case studies; note that they only require the prior knowledge of the Cu, Al and separator layer thicknesses, which are assumed to be raw production parameters and easily measurable.

\subsection{Identifying number of electrode layers}\label{sec:layer_number}

We can estimate the total number of electrode layers (i.e. resonant elements) inside a pouch cell from the ultrasonic resonance. When the cell is thin and the resonance is formed throughout its thickness, the number of layers is simply equal to the number of resonant peaks. When the cell is thick and the resonance is only formed on the first 20 layers or so, this number can be estimated by obtaining two phase shift values (physically equivalent to TOF) of the propagated wave from the ultrasonic measurements: the phase shift of a resonance element, and that of the whole cell. The number of layers is simply estimated by dividing the latter with the former. 

The first is the phase shift $\phi_\mathrm{elem}$ of a wave travelling a round trip through the space of a resonant element, which consists of one Cu and one Al current collector and a combined anode-separator-cathode layer. The known thicknesses of the metal layers enable the phase shift $\phi_\mathrm{e}$ of the combined layer to be obtained from experimental resonance via Eq. \ref{eq:resonance}. However, one important nuance here is that the phase shifts of the metal layers $\phi_\mathrm{tc}^*$ and $\phi_\mathrm{ta}^*$ (emphasised with asterisks) that contribute to $\phi_\mathrm{elem}$ should be calculated from direct transmission through the layers - unlike the phase shifts in Eq. \ref{eq:resonance}, which implicitly include multiple internal reverberations. This is because the back-wall echo (for determining $\phi_\mathrm{total}$ below) is the first-arrival signal, which requires the wave to go straight through each layer to satisfy the shortest travel time. Considering the actual wave paths in individual layers of the the Kokam cell, the phase change in a resonant element is given as $\phi_\mathrm{elem}=\phi_\mathrm{tc}^*+2\phi_\mathrm{e}+\phi_\mathrm{ta}^*=-5.70$ rad.

\begin{figure}[ht]
    \centering
    \includegraphics[width=0.9\linewidth]{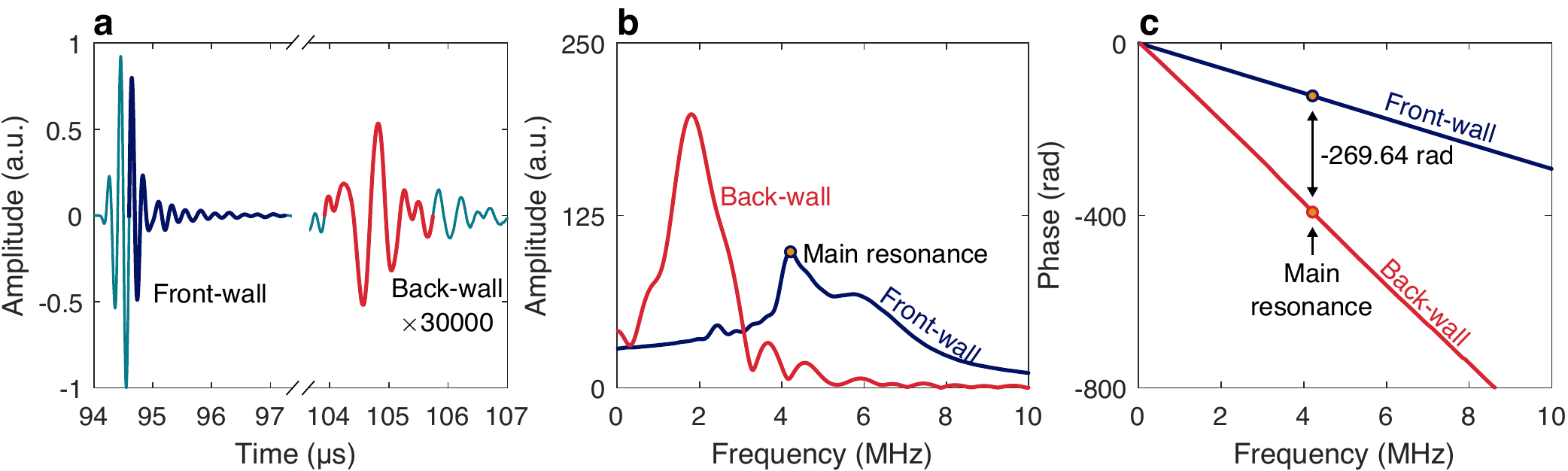}
    \caption{Front- and back-wall reflections from the Kokam 7.5 Ah cell. \textbf{a} shows the time-domain signals acquired with an immersion setup using a 5 MHz probe placed 70.5 mm apart from the cell. Here the back-wall echo was acquired and amplified from a battery-air interface (battery-water reflection was not strong enough) created by water-sealing an air-filled container at the back-wall. \textbf{b} and \textbf{c} present the respective amplitude and unwrapped phase spectra of the front- and back-wall echoes.}
    \label{fig:layer_number}
\end{figure}

The second is the total phase shift $\phi_\mathrm{total}$ of the wave travelling a round trip through the whole cell. This can be obtained from the reflections from the front- and back-wall of the cell. Such signals are plotted in Fig. \ref{fig:layer_number}a for the Kokam cell, which were acquired using the pulse-echo setup, with the 5 MHz probe placed 70.5 mm away from the cell. Despite noticeably different amplitude spectra in Fig. \ref{fig:layer_number}b, the phase spectra of these two signals in Fig. \ref{fig:layer_number}c show good linearity and estimate the phase difference at the main resonance frequency to be $\phi=-269.64$ rad. This then needs to exclude the phase delays caused by the two surface packaging layers of the cell, and the phase reversal at the front-wall (due to wave incident from low-impedance water to high-impedance cell), which eventually gives the total round-trip phase change as $\phi_\mathrm{total}=-270.94$ rad.

The number $N$ of resonant elements (thus the number of electrode layer pairs) of the cell, can be now calculated simply by $N=\phi_\mathrm{total}/\phi_\mathrm{elem}=47.51$. This number is very close to the actual number of 48 as counted by destructive tests \cite{Ecker2015,Hales2019}. Moreover, excellent agreement of $N$ was obtained from further contact tests on the same cell in Appendix \ref{appdx:layer_number}, proving the robustness and reproducibility of the estimation.

\subsection{Estimating average electrode thicknesses}

Here the ultrasonic resonance is used to estimate another average property of the examined cell, i.e., the average thicknesses of its anode and cathode layers, $d_\mathrm{an}$ and $d_\mathrm{ca}$. This is achieved by formulating and solving two independent equations, respectively from the phase (or TOF) and thickness aspects, for $d_\mathrm{an}$ and $d_\mathrm{ca}$. 

The first equation is based on the phase shift aspect. From the main resonance formation condition in Eq. \ref{eq:resonance}, the phase shift $\phi_\mathrm{e}$ in the combined anode-separator-cathode layer, at the resonant frequency of $f_\mathrm{r}\approx4.17$ MHz, has been determined from time traces in the previous subsection. By breaking $\phi_\mathrm{e}$ down to the individual layers, and assuming that the wave speeds in the electrodes are known from the slurry model, we arrive at the first equation:
\begin{linenomath*}
\begin{equation}
    \phi_\mathrm{e}=-2\pi f_\mathrm{r} (d_\mathrm{an}/c_\mathrm{an}+d_\mathrm{ca}/c_\mathrm{ca}+d_\mathrm{s}/c_\mathrm{s})=-2.78\,\mathrm{rad},
\end{equation}
\end{linenomath*}
where only $d_\mathrm{an}$ and $d_\mathrm{ca}$ are unknown, and all the other parameters are given in Table \ref{tab:properties}.

The second equation is simply constructed from the thickness relationships. The whole cell was measured to be 7.26 mm thick, while the thicknesses of all packaging, separator, Cu and Al layers are summed up to be 1.90 mm based on the measured values in Table \ref{tab:properties}. Subtracting the latter from the former, we obtain the total thickness of the 48 layers of anode and cathode, written as:
\begin{linenomath*}
\begin{equation}
    48(d_\mathrm{an}+d_\mathrm{ca})=7.26-1.90=5.36\,\mathrm{mm}.
\end{equation}
\end{linenomath*}

Solving these two equations gives the average thicknesses of $d_\mathrm{an}=\SI{69.18}{\micro\meter}$ and $d_\mathrm{ca}=\SI{42.49}{\micro\meter}$ for a single anode and cathode layer respectively. As a way of verification, we calculated the nominal positive/negative electrode capacity ratio of the cell based on the thicknesses estimated here and those destructively measured \cite{Ecker2015}, under an identical assumption that the anode and cathode layers have the same proportion of active materials (with theoretical capacities of $q_\mathrm{an}=372\,\mathrm{mAh/g}$ or $710\,\mathrm{mAh/cm^3}$ and $q_\mathrm{ca}=274\,\mathrm{mAh/g}$ or $1143\,\mathrm{mAh/cm^3}$ \cite{Nitta2015}). The results are 1.01 calculated here versus 0.84 measured destructively, with the former falling well into the typical design range of 1.0-1.2 \cite{Wu2019}. Therefore, we suspect that our estimations could be more accurate than destructive measurements (though note that the estimation is sensitive to the accuracy of wave speeds in electrodes), especially considering that the electrodes, which are compressible materials, could expand considerably and unevenly when the battery is dismantled.

\subsection{Constructing image of internal layers}

\begin{figure}[ht]
	\centering
	\includegraphics[width=\linewidth]{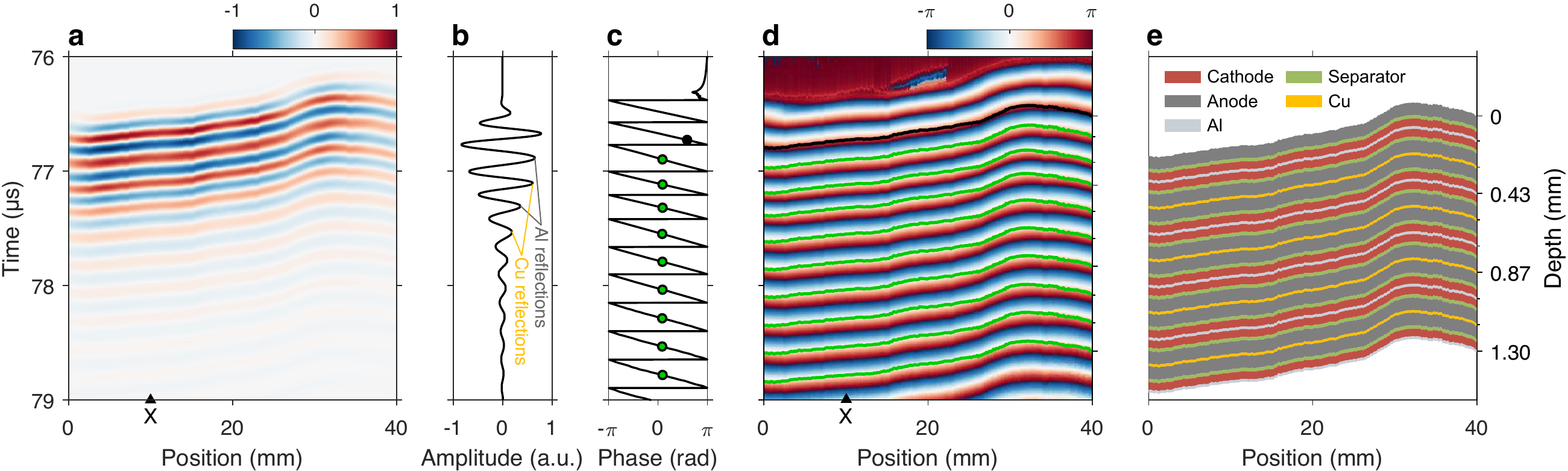}
	\caption{Constructing the image of battery internal layers using experimental signals. \textbf{a} shows a B-scan of ultrasonic signals over a range of 40 mm on the Kokam 7.5 Ah cell; the signals were acquired with the immersion setup shown in Fig. \ref{fig:experiments}e using a 5 MHz probe; the probe has a distance of 28.5 mm to the cell and each signal is the second wave packet of the originally acquired signal. \textbf{b} demonstrates the signal at point X as annotated in \textbf{a}. The colours in \textbf{a} represent the amplitudes in \textbf{b}. Figure \textbf{c} plots the respective instantaneous phase at point X; the black point corresponds to the value representing the front surface of the cell; the green points have a phase difference of $-\pi/2$ to the black point and represent the metal layers within the cell. \textbf{d} displays a stack of instantaneous phases for the scanned positions in \textbf{a}; the cell front surface is identified by a black line and the internal metal layers by green lines. \textbf{e} illustrates the reconstructed layer profile of the cell.}
	\label{fig:imaging_exp}
\end{figure}

Now we utilise the layer-resolved characteristic of the resonant signal to construct an image of the cell's internal layers. For this purpose, we performed a line scan (B-scan) of the Kokam 7.5 Ah cell using the immersion setup, which allows the front surface of the cell to be captured (alternatively, contact tests with a delay line can achieve the same purpose). We used the 5 MHz immersion probe, which was placed 28.5 mm away from the cell and scanned over a 40 mm line in the middle of the cell, with a step size of 0.125 mm. Local spot inspection (A-scan) signals were recorded at each scanning step, delivering a sequence of 321 individual signals along the line. All the results are plotted as a B-scan map in Fig. \ref{fig:imaging_exp}a, where each A-scan signal (e.g. \ref{fig:imaging_exp}b) was plotted as a vertical line, with its varying amplitude indicated by the changing colours. Thus, Fig. \ref{fig:imaging_exp}a illustrates the arrival time variations across different scanned locations due to the spatially-varied layer profile of the cell. 

Fig. \ref{fig:imaging_exp}b demonstrates well-formed resonance, with gradually decaying amplitude. To use the time trace for layer reconstruction, the front surface and internal metal layers of the cell are located from each testing position. This is achieved through the analytical signal method, as outlined by Smith et al. \cite{Smith2018}: if performed the Hilbert transform, the experimental time-domain signal is transformed into a complex analytic signal, whose instantaneous amplitude, phase and frequency of the analytical signal all have direct correspondence to the locations of the reflection interfaces. The instantaneous phase curve of the A-scan signal in Fig. \ref{fig:imaging_exp}b, given in Fig. \ref{fig:imaging_exp}c, is used below as an example, but in general, the instantaneous amplitudes and frequencies can also be used 
separately or collectively \cite{Smith2018}. 

Following Smith et al\cite{Smith2018}, the front surface of the cell should be located at the peak of instantaneous amplitude, shown as the top solid point in Fig. \ref{fig:imaging_exp}c, while the internal reflecting metal layers correspond to points with a phase difference of $-\pi/2$ from the front surface, indicated by the subsequent points in the figure. By doing so, the surface profile of the cell over the scanning line is obtained as the top black line in Fig. \ref{fig:imaging_exp}d, and the internal metal layer profiles as the subsequent green lines. Furthermore, the different reflection amplitudes shown in Fig. \ref{fig:imaging_exp}b (also highlighted in Fig. \ref{fig:experiments}c) allow the Al and Cu layers to be differentiated. 

With the metal layers located, the next step is to fill the anode, cathode and separator layers in between. The separator thickness is known (Table \ref{tab:properties}), while the electrode layers are based on the average thicknesses estimated from the central resonant frequency in the preceding subsection. Note that due to localised curvature (e.g. waves in non-normal directions travel longer distances) and uneven thicknesses of the layers, the peak-to-peak gaps in the time-domain signal may deviate slightly from the resonance estimation. In that case, it is assumed that the ratio between the anode and cathode remains the same as estimated in the previous subsection, and both thicknesses are scaled linearly to fit the distance. The end result of this image construction is plotted in Fig. \ref{fig:imaging_exp}e, where the individual internal layers of the cell are successfully identified and tracked. For clarity, it only showcases several shallow metal layers; however, the signals can reliably deliver structural information of the first 10-15 peaks from one side (further peaks are limited by lower amplitudes, but inspections can be performed on both sides).

\subsection{SOC monitoring with layer resolution}

The electrochemical reactions during charging and discharging of a battery modify the key physical properties of the electrode layers, including elastic constants, density and thickness, which affect, and thus could be detected by, the ultrasonic resonance, potentially with single-layer resolution.

\begin{figure}[ht]
	\centering
	\includegraphics[width=\linewidth]{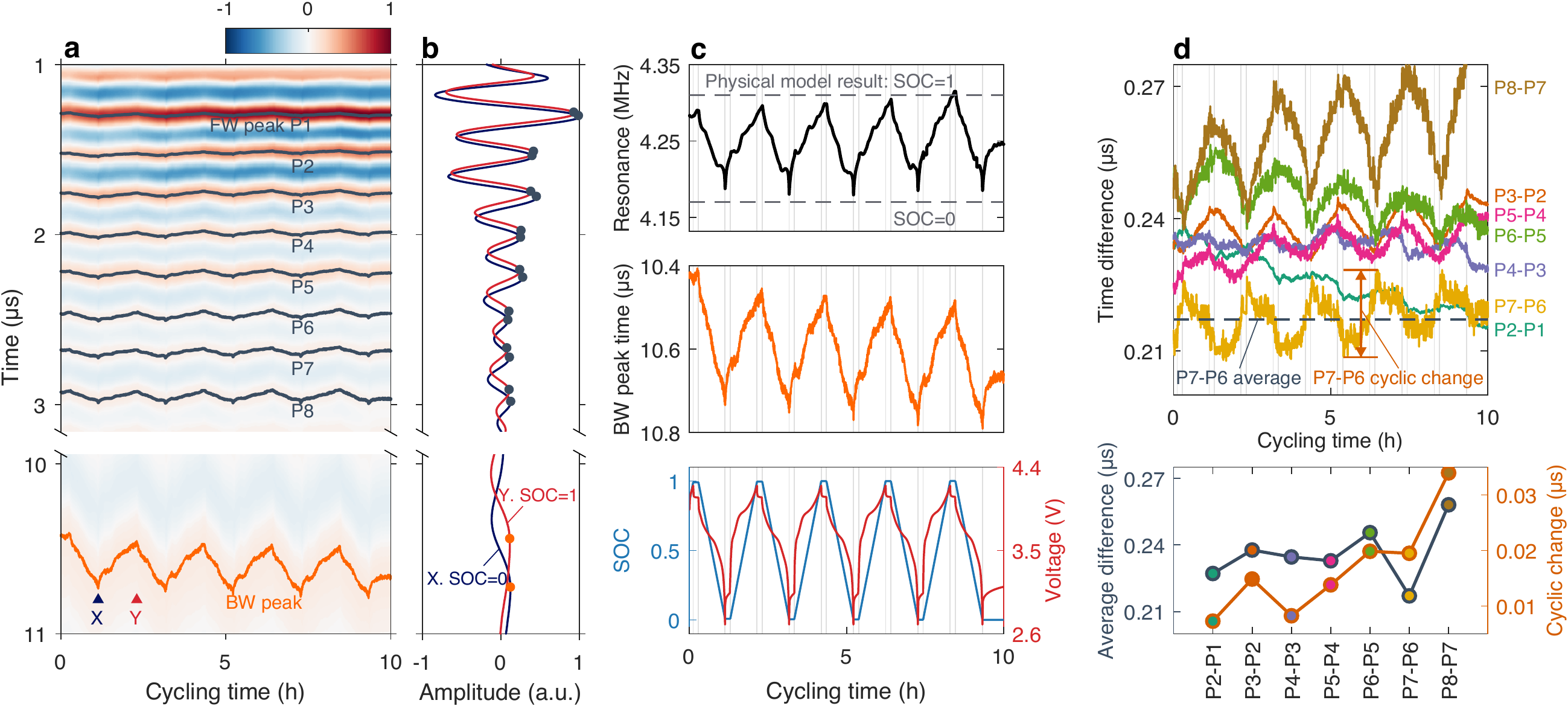}
	\caption{SOC monitoring using ultrasonic signals. \textbf{a} shows a map of ultrasonic signals monitored every 10 seconds using a 5 MHz contact probe on the Kokam 7.5 Ah cell during the charging and discharging experiment. The upper half is for reflections by the internal layers while the lower half for the back-wall reflection. The dark blue (P1 to P8) and orange lines highlight the peaks of the signals. The time points at X and Y correspond to SOC=0 and SOC=1, respectively.  \textbf{b} displays the respective signals at X and Y. \textbf{c} compares the cyclic resonant frequency (top) and back-wall signal peak (middle) with the SOC and voltage curves (bottom). The top figure of \textbf{d} shows the time differences between neighbouring front-wall peaks versus cycling time, while the bottom figure displays the respective averages (in blue line) and cyclic changes (in orange) for all the individual peak pairs.}
	\label{fig:soc_monitoring}
\end{figure}

To investigate this, we performed 5 full charge-discharge cycles on the Kokam 7.5 Ah cell using a Biologic BCS-815 battery cycler. The cell was charged and discharged with a constant current at a rate of 1C, with upper and lower voltage limits of 4.2 and 2.7 V. A 10 minute rest period was applied after each charge or discharge process, with the cell housed in a Binder thermal chamber at 25 \textdegree C throughout. The cell SOC was calculated via coulomb counting (sample interval of 2 ms), normalising the charge passed by the maximum capacity obtained during the experiment. Over the whole period, the battery was monitored using the 5 MHz contact probe (V109-RM, Olympus), which was pressed against the cell surface by a spring mechanism to maintain consistent contact. Ultrasonic signals were recorded every 10 seconds, thus the 10-hour test (2 hours per full cycle) led to 3600 recordings, all plotted together as a map changing with time in Fig. \ref{fig:soc_monitoring}a. The upper half of the map highlights the resonance, exemplified by the dark blue lines tracking the first few peaks; while the lower half tracks the back-wall echoes, demonstrated by the orange line. We emphasise that the latter delivers similar information to the through-transmission configurations adopted in prior studies \cite{hsieh2015electrochemical,davies2017state,chang2020measuring}.

In Fig. \ref{fig:soc_monitoring}a, the greatest contrast of the ultrasonic signals happen between SOC=0 and 1, e.g. at the moments X and Y in the figure, and their time traces are compared in Fig. \ref{fig:soc_monitoring}b. The signal at SOC=1 has a noticeably shorter propagation time, as if linearly compressed like a spring, compared to SOC=0. This is due to increases in overall wave speeds through the electrode layers with increasing SOC (Table \ref{tab:properties}).

As a result, the central resonant frequency of the time traces, which implicitly accounts for all the analysed peaks through Fourier transform, has similar sensitivity as the back-wall echo. This is illustrated in Fig. \ref{fig:soc_monitoring}c, where both the resonant frequency and the TOF of the back-wall reflection display good correlation with coulomb counting and, in particular, voltage estimations of SOC. Furthermore, the resonant frequencies at SOC=0 and SOC=1 predicted by our physical model, marked by the dashed lines in Fig \ref{fig:soc_monitoring}c, match almost perfectly with experimental results, further validating our analytical model.

More excitingly, the resonance method allows monitoring of layer-by-layer property changes during cycling. Since the resonance peaks track the positions of metal layers, the gap between two neighbouring peaks gives information about a single resonant element, i.e. one anode and one cathode layer. The cyclic behaviours of the first seven elements are plotted in Fig. \ref{fig:soc_monitoring}d, and immediately noticeable are two prominent features. Firstly, the curves are slightly shifted vertically from each other (plotted below in blue line as 'Average difference'). This indicates different travelling times through the elements even without charging, likely caused by small variations of layer thicknesses. Secondly, the cyclic changes (plotted in orange below) varies considerably across layers. For example, the TOF changes for P8-P7 are almost three times as much as P2-P1, and they appear to grow more pronounced deeper into the battery. These observations strongly indicate that the state changes during cycling are happening heterogeneously across layers. Indeed, similar results were delivered by electro-thermal models \cite{LI2021229594, zhao2018modeling}, which also revealed that the variations were largely due to differences in resistance for each layer, and the positive feedback with current and temperature. Even though the fact that such heterogeneity exists in the battery bulk is experimentally observable, e.g. by the different SOCs estimated by coulomb counting (which gives the bulk SOC) and voltage (influenced by specific parts of the cell which are at different SOCs than the bulk), there is currently no way to estimate distributed SOC levels from normal electrochemical data. Therefore, we believe that the layer-resolved SOC monitoring capability of the resonance method (which will require dedicated studies to fully develop) offers a much-need and powerful addition to the existing battery management tools.

\section{Discussions and conclusion}\label{sec:conclusions}

In this paper, we have established an advanced methodology for characterising the layer properties and states of LIBs from ultrasonic resonance. Robust experimental acquisitions of resonant signals were achieved, and a comprehensive theoretical model was established by analysing the wave physics in individual layers of an LIB (a notable contribution is treating electrodes as dense slurries) and the interference of their reflections to form the resonance. We demonstrated high levels of accuracy of the developed approach in comparison to experimental results (e.g. predicted resonant frequency error $\sim$1\% at different SOCs), and showcased its efficacy in quantitatively characterising the number of electrode layers in the cell, the average thickness of the anode and cathode layers, the images of internal structure and SOC resolved up to layer level during electrochemical cycling.

The proofs of principle established in this paper open the door for a range of exciting possibilities for further, more in-depth research, as well as real-world applications. Specifically, the central resonant frequency enables accurate and reliable inversion of battery SOC, and the variation of TOF between resonant peaks in the depth direction may help understand the in-operando temperature gradients or heterogeneity in electrochemical reactions. The sensitivity of acoustic behaviours to material changes may facilitate the monitoring of battery SOH by quantitative full-waveform studies. For example, lithium plating could cause the resonance to shift to a higher frequency, while degradation-associated cracks, dislocations and porosity changes of the particles \cite{robinson2020identifying} may result in higher amplitude attenuation of the resonant peaks. Volume expansion in electrode layers could also be determined from the resonance, which may be of particular interest for next-generation, silicon-based anode materials. Many of these investigations could be further transported to the characterisation of cylindrical batteries which occupy a large share of the market.

In addition, opportunities could emerge in developing new experimental techniques for resonant signal acquisition. The bulky and intrusive probes used in this paper could be replaced with thin and flexible micro-electromechanical ones (already used on e.g. mobile phones), for permanently-installed monitoring, or as an addition to the battery management system. Non-contact, air-coupled probes could be used for in-production (e.g. for electrode porosity) or in-service monitoring, to avoid using water or gel as couplant. These developments can be explored in parallel with the research topics outlined above, and can easily incorporate the latest theoretical progress. Moreover, a growing set of experimental data will benefit both the experimental and theoretical aspects of the work, since statistical analyses and machine learning tools can be employed to enhance the accuracy of material properties and aid physics-based interpretations, thus constantly improving the reliability of the methodology.

\section*{Acknowledgements}
B.L. gratefully acknowledges the generous support from the Imperial College Research Fellowship Scheme, M.H. the Non-Destructive Evaluation Group at Imperial College, and N.K. Innovate UK through the WIZer project (Grant No. 104427). We would like to thank Prof Robert Smith of Bristol University for helpful discussions on the resonance method, and Prof Peter Cawley of Imperial College for reviewing the manuscript.

\appendix
\section*{Appendices}
\counterwithin{figure}{section}
\counterwithin{table}{section}
\counterwithin{equation}{section}

\section{Observation of ultrasonic resonance from two more LIBs}\label{appdx:experiments}
\begin{figure}[!ht]
    \centering
    \includegraphics[width=0.5\linewidth]{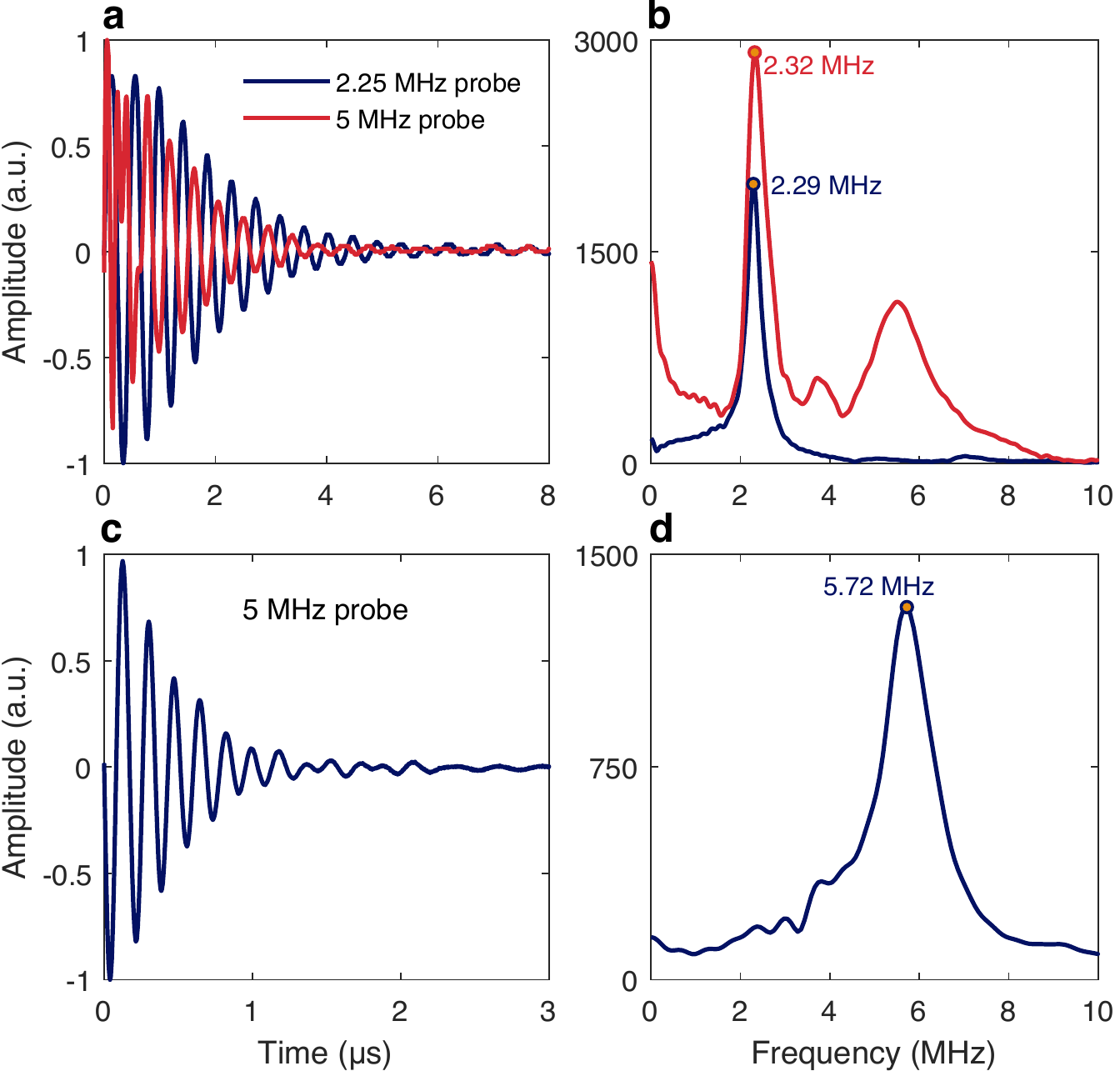}
    \caption{Observation of ultrasonic resonance from two more LIBs. \textbf{a} shows the time-domain signals of 2.25 MHz and 5 MHz contact probes from a 210 mAh battery cell (PL-651628-2C, AA Portable Power Corp.), while \textbf{b} displays the respective amplitude spectra. Similarly, \textbf{c} illustrates the time-domain signals of a 5 MHz contact probe from a 5 Ah cell (SLPB11543140H5, Kokam) and \textbf{d} presents the respective amplitude spectrum.}
    \label{fig:more_experiments}
\end{figure}

We conducted ultrasonic contact tests on two more LIB pouch cells, including a 210 mAh cell (PL-651628-2C, AA Portable Power Corp.) and a Kokam 5 Ah cell (SLPB11543140H5). The acquired time- and frequency-domain results are shown in Fig. \ref{fig:more_experiments}a and b for the former cell, and those for the latter are provided in c and d. These results exhibit strong resonance similar to those in Fig. \ref{fig:experiments} of the main text.

\section{Modelling separators using the Biot model}\label{appdx:separator}

The separators in a LIB are usually porous solids filled with liquid electrolytes \cite{Martinez-Cisneros2016}, acting as an ionically-conductive physical barrier between two electrodes. The Biot model \cite{biot1,biot2} is thus needed to describe the propagation of ultrasonic waves within such a fluid-saturated porous separator. This model predicts the existence of three waves, one shear and two longitudinal. The shear wave is not involved since we consider longitudinal waves only while the slow longitudinal wave barely contributes to the detected ultrasonic signal due to its diffuse nature and excessive attenuation, so our concern is with the fast longitudinal wave only. This fast wave can be treated as independent of frequency in the frequency range utilised in this work, which is lower than the transition frequency $f_\mathrm{t}=\pi\mu/(4\rho_\mathrm{L}d^2)=10\,\mathrm{MHz}$ \cite{biot1}. This transition frequency estimation was based on the material parameters of the Kokam 7.5 Ah battery, for which we used a large pore diameter \cite{Martinez-Cisneros2016} $d=0.5\,\SI{}{\micro\meter}$ for the porous polypropylene solid for a cautious estimation, while used the viscosity \cite{Gor2014} $\mu=4.2\,\mathrm{mPa\cdot s}$ and density \cite{gold2017probing} $\rho_\mathrm{L}=1270\,\mathrm{kg/m^3}$ for the liquid $\mathrm{LiPF_6}$ electrolyte \cite{Ecker2015}. In this low-frequency range, the wave speed is given by \cite{biot1} $c=\sqrt{H/\rho}$. The effective density of the medium is related to the porosity $\nu$ and the densities $\rho_\mathrm{S}$ and $\rho_\mathrm{L}$ of the solid and liquid by $\rho=(1-\nu)\rho_\mathrm{S}+\nu\rho_\mathrm{L}$. The subscript S here refers to the homogeneous solid without pores, while the subscript P below denotes the porous solid. $H=A+2N+R+2Q$ is the effective longitudinal modulus of the medium, with $A$ and $N$ being the Lam\'e constants, $R$ the pressure required for forcing a certain volume of the liquid into the medium whilst the total volume remains constant, and $Q$ the coupling of volume change between the solid and liquid. These four parameters are given by \cite{Biot1957,Jocker2009}:
\begin{linenomath*}
\begin{equation}\small
        N=G_\mathrm{P},\; A=K_\mathrm{P}-2N/3+K_\mathrm{L}(1-\nu-K_\mathrm{P}/K_\mathrm{S})^2/\nu_\mathrm{eff},\; Q=\nu K_\mathrm{L}(1-\nu-K_\mathrm{P}/K_\mathrm{S})/\nu_\mathrm{eff},\; R=\nu^2K_\mathrm{L}/\nu_\mathrm{eff},
\end{equation}
\end{linenomath*}
where $K_\mathrm{S}$ and $K_\mathrm{L}$ are the bulk moduli of the solid and the liquid respectively. $\nu_\mathrm{eff}=\nu+K_\mathrm{L}/K_\mathrm{S}(1-\nu-K_\mathrm{P}/K_\mathrm{S})$ is an effective porosity. $K_\mathrm{P}$ and $G_\mathrm{P}$ are the bulk and shear moduli of the porous solid, which are related to the porosity and the properties of the homogeneous solid; we determine these two parameters using the Mori-Tanaka mean field theory \cite{Zhao1989}:
\begin{linenomath*}
\begin{equation}\small
        K_\mathrm{P}=4G_\mathrm{S}K_\mathrm{S}(1-\nu)/(4G_\mathrm{S}+3\nu K_\mathrm{S}),\; G_\mathrm{P}=G_\mathrm{S}(8G_\mathrm{S}+9K_\mathrm{S})(1-\nu)/(8G_\mathrm{S}+9K_\mathrm{S}+6(2G_\mathrm{S}+K_\mathrm{S})\nu).
\end{equation}
\end{linenomath*}

By using the material properties in Table \ref{tab:separator}, we calculated the wave speed and density of the separator for the Kokam 7.5 Ah battery cell and the results are given in Table \ref{tab:properties}.

\begin{table}[ht]
    \centering
    \caption{Parameters for calculating the material properties of liquid-filled porous separator using the Biot model \cite{biot1}. The solid phase S is polypropylene and the liquid phase L is $\mathrm{LiPF_6}$ \cite{Ecker2015}.}
    \begin{tabular}{lccccc}
        \hline
        $\nu$ & $K_\mathrm{S}$ (GPa) & $G_\mathrm{S}$ (GPa) & $\rho_\mathrm{S}\,\mathrm{(kg/m^3)}$  & $K_\mathrm{L}$ (GPa) & $\rho_\mathrm{L}\,\mathrm{(kg/m^3)}$\\
        \hline
        0.508 \cite{Ecker2015} & 2.2 \cite{Bedoui2004} & 0.3 \cite{Bedoui2004} & 850 \cite{Bedoui2004} & 1 \cite{gold2017probing} & 1270 \cite{gold2017probing}\\
        \hline
    \end{tabular}
    \label{tab:separator}
\end{table}

\section{Estimating electrode properties with the slurry model}\label{appdx:slurry}

We have proposed in section \ref{sec:layers} that the battery electrodes should be modelled as dense slurries. The acoustic model of a slurry is well-established, and the wave speed can be calculated from \cite{ATKINSON1992577}:
\begin{linenomath*}
\begin{equation} \label{eq:slurry1}
    c = \sqrt{\frac{K}{\rho}},
\end{equation}
\end{linenomath*}
with the bulk modulus $K$ and the density $\rho$ of the slurry given by:
\begin{linenomath*}
\begin{equation} \label{eq:slurry2}
    \frac{1}{K} = \frac{ \nu }{K_\mathrm{S}} + \frac{1-\nu}{K_\mathrm{L}}, \; \rho = \nu \rho_\mathrm{S} + (1-\nu)\rho_\mathrm{L},
\end{equation}
\end{linenomath*}
where the subscripts of S and L respectively refer to the solid and liquid phases of the slurry, and $\nu$ the volume fraction of the solid phase, which is an important factor and needs to be pre-determined for the calculations. For the Kokam 7.5 Ah cell, we used the volume fractions of 0.773 and 0.811 for anode and cathode, which were obtained from the porosity data as determined by mercury porosimetry \cite{Ecker2015}. By substituting the material properties listed in Table \ref{tab:electrodes} into Eqn. \ref{eq:slurry2}, we obtain the bulk moduli and densities of the electrodes as shown in the same table. Following this, the respective wave speeds are calculated by Eqn. \ref{eq:slurry1} and given in Table \ref{tab:properties}.

\begin{table}[ht]
    \centering
    \begin{threeparttable}
    \caption{Electrode properties of Kokam 7.5 Ah pouch cell: original solid and liquid phases properties and predicted effective properties using a slurry model \cite{ATKINSON1992577}. Units: bulk modulus $K$ (GPa), density $\rho\,\mathrm{(kg/m^3)}$, solid phase volume fraction $\nu$ (dimensionless).}
    \label{tab:electrodes}
    \begin{tabular}{lccccccccc}
        \hline
        Layer & Solid \cite{davies2017state} & $K_\mathrm{S}$ \cite{davies2017state} & $\rho_\mathrm{S}$ \cite{davies2017state} & $\nu$ & Liquid \cite{Ecker2015} & $K_\mathrm{L}$ \cite{gold2017probing} & $\rho_\mathrm{L}$ \cite{gold2017probing} & $K$ & $\rho$ \\
        \hline
        Anode (SOC=0)   & $\mathrm{C}_6$ & 28.8  & 2260 & $0.773^\ddag$ (0.671 \cite{Ecker2015}) & \multirow{4}{*}{$\mathrm{LiPF_6}$} & \multirow{4}{*}{1} & \multirow{4}{*}{1270} & 3.43 & 1909 \\
        Cathode (SOC=0) & $\mathrm{Li_{0.95}CoO_2}^\dag$ & 88.9 & 4860 & $0.811^\ddag$ (0.704 \cite{Ecker2015}) &&&& 4.98 & 4172  \\
        Anode (SOC=1)  & $\mathrm{Li_{0.85}C_6}$ & 67.8  & 2210 & $0.773^\ddag$ (0.671 \cite{Ecker2015}) &&&& 4.15 & 1994  \\
        Cathode (SOC=1) & $\mathrm{Li_{0.5}CoO_2}^\dag$ & 82.4 & 4460 & $0.811^\ddag$ (0.704 \cite{Ecker2015}) &&&& 4.96 & 3848  \\
        \hline
    \end{tabular}
    \begin{tablenotes}
        \item[$^\dag$] The cell has an as-built cathode of $\mathrm{Li(Ni_{0.4}Co_{0.6})O_2}$ \cite{Ecker2015,Hales2019} containing nickel and cobalt; here we treat nickel and cobalt as a single composition of cobalt because they have similar properties.
        \item[$^\ddag$] The volume fractions were obtained by scaling the destructively measured values in parentheses in order for the total thickness of the cell to match the actual value; the scaling was performed under the condition of volume conservation for the solid phase.
    \end{tablenotes}
    \end{threeparttable}
\end{table}

\section{Full-waveform modelling using the transfer matrix method}\label{appdx:full_waveform}

The general idea of the full-waveform modelling is shown in Fig. \ref{fig:resonance}e. Instead of the individual time-domain reflections (which contain a broad frequency bandwidth) from different layers and matching up their phases, the analysis here is carried out in the frequency domain for individual frequencies and then transformed back to the time domain. When the multilayered medium is subject to a monochromatic longitudinal wave $u_0^+(x)=a_0^+e^{-ik_0x}$, two wave components would arise in an arbitrary layer $n$, propagating in the forward and backward directions. The total displacement at $x$ in the layer is given by the summation of the two components as $u_n(x)=a_n^+e^{-ik_n(x-x_{n-1})}+a_n^-e^{ik_n(x-x_{n-1})}$. $x_{n-1}$ and $x_{n}$ are the coordinates of the two interfaces. $a$ carries both the amplitude and phase information of the wave. $k_n=\omega/c_n$ is the wave number; $\omega$ is the angular frequency of the incident wave and $c_n$ is the longitudinal wave speed of the layer. For simplicity, the time harmonic $e^{-i\omega t}$ is neglected in the displacement expressions.

To calculate the actual wave $u_n(x)$, we need to determine the wave amplitudes $a_n^+$ and $a_n^-$. The way to achieve this is to use the fundamental fact that stress and velocity must be continuous across any boundary to satisfy the equilibrium and compatibility conditions. In any layer $n$, the stress and velocity are related to the displacement by:
\begin{linenomath*}
\begin{equation}\label{eq:waves_relations}
	\begin{aligned}
		\sigma_n(x)&=M_n\frac{du_n(x)}{dx}=-i\omega Z_n[a_n^+e^{-ik_n(x-x_{n-1})}-a_n^-e^{ik_n(x-x_{n-1})}],\\
		v_n(x)&=\frac{du_n(x)}{dt}=-i\omega[a_n^+e^{-ik_n(x-x_{n-1})}+a_n^-e^{ik_n(x-x_{n-1})}].
	\end{aligned}
\end{equation}
\end{linenomath*}
where $M_n$ is the longitudinal modulus of the layer and $Z_n=M_n/c_n=\rho_nc_n$ is the acoustic impedance. For fluid, stress corresponds to pressure and longitudinal modulus to bulk modulus. Therefore, the continuity of stress and velocity on each boundary delivers two equations for the wave amplitudes. For instance, the equations are as follows for boundaries 0 (bordering layers $n=0$ and $n=1$) and 1 (bordering $n=1$ and $n=2$):
\begin{linenomath*}
\begin{equation}
    \begin{aligned}
    \sigma_0(x_0)=\sigma_1(x_0): & \quad Z_0a_0^+ - Z_0a_0^- - Z_1a_1^+ + Z_1a_1^-=0\\
    v_0(x_0)=v_1(x_0): & \quad a_0^+ + a_0^- - a_1^+ - a_1^-=0\\
    \sigma_1(x_1)=\sigma_2(x_1): & \quad Z_1e^{-ik_1d_1}a_1^+ - Z_1e^{ik_1d_1}a_1^- - Z_2a_2^+ + Z_2a_2^-=0\\
    v_1(x_1)=v_2(x_1): & \quad e^{-ik_1d_1}a_1^+ + e^{ik_1d_1}a_1^- - a_2^+ - a_2^-=0\\
    \end{aligned}
\end{equation}
\end{linenomath*}
Altogether, a system of $2(N+1)$ equations can be constructed by using the continuity conditions for all $N+1$ boundaries. Practically, the incident amplitude $a_0^+$ is given beforehand, and $a_{N+1}^-$ is zero because no wave comes into the layers from the back face. As a result, the number of equations is the same as the number of unknowns, which are $a_0^-$, $a_n^+$ and $a_n^-$ ($n=1,2,\ldots,N$), and $a_{N+1}^+$; therefore, solving the equation system gives the solutions for all unknown amplitudes.

Since we use the pulse-echo setup that sends and receives signals with the same probe, we are only interested in the solution for $a_0^-$ and, in particular, its ratio to the incident amplitude as characterised by the total reflection coefficient $R=a_0^-/a_0^+$. For this reason, we have formulated a very computationally-efficient transfer matrix scheme that solves only for $R$ recursively. To achieve this, we write the stress and velocity at the two interfaces $x_{n-1}$ and $x_{n}$ of the layer in matrix form (based on Eq. \ref{eq:waves_relations}):
\begin{linenomath*}
\begin{equation}\label{eq:waves_left}
	\begin{bmatrix} \sigma(x_{n-1})\\ v(x_{n-1}) \end{bmatrix}_{n}=-i\omega \begin{bmatrix} -Z_n&Z_n \\ 1&1 \end{bmatrix}	\begin{bmatrix} a_n^-\\ a_n^+ \end{bmatrix},
\end{equation}
\end{linenomath*}
and
\begin{linenomath*}
\begin{equation}\label{eq:waves_right}
	\begin{bmatrix} \sigma(x_{n})\\ v(x_{n}) \end{bmatrix}_{n}=-i\omega \begin{bmatrix} -Z_ne^{ik_nd_n}&Z_ne^{-ik_nd_n} \\ e^{ik_nd_n}&e^{-ik_nd_n} \end{bmatrix}	\begin{bmatrix} a_n^-\\ a_n^+ \end{bmatrix},
\end{equation}
\end{linenomath*}
where $d_n$ is the layer thickness. Solving Eq. \ref{eq:waves_right} for $[a_n^-,a_n^+]^\mathrm{T}$ and substituting the solution into Eq. \ref{eq:waves_left} would lead to the transfer matrix equation:
\begin{linenomath*}
\begin{equation}\label{eq:waves_transfer}
	\begin{bmatrix} \sigma(x_{n-1})\\ v(x_{n-1}) \end{bmatrix}_{n}=\begin{bmatrix} \cos(k_nd_n)&iZ_n\sin(k_nd_n) \\ i\sin(k_nd_n)/Z_n&\cos(k_nd_n) \end{bmatrix} \begin{bmatrix} \sigma(x_{n})\\ v(x_{n}) \end{bmatrix}_{n},
\end{equation}
\end{linenomath*}
which relates (transfers) the stress and velocity of interface $x_{n}$ to those of $x_{n-1}$.

In the bounding media $N+1$, owing to the absence of backward propagating wave ($a_{N+1}^-=0$), it follows from Eq. \ref{eq:waves_left} that $\sigma_{N+1}(x_N)/v_{N+1}(x_N)=Z_{N+1}$ at the interface $N$. Considering the continuity of stress and velocity at the interface $N$, we have $\sigma_{N}(x_N)/v_{N}(x_N)=\sigma_{N+1}(x_N)/v_{N+1}(x_N)=Z_{N+1}$. When the stress and velocity transfer to the interface $N-1$, an effective impedance for the combination of layers $N$ and $N+1$ would be obtained from Eq. \ref{eq:waves_transfer}, given by:
\begin{linenomath*}
\begin{equation}
	Z_N^\mathrm{eff}=\frac{\sigma_{N}(x_{N-1})}{v_{N}(x_{N-1})}=Z_N\frac{Z_{N+1}\cos(k_Nd_N)+iZ_N\sin(k_Nd_N)}{iZ_{N+1}\sin(k_Nd_N)+Z_N\cos(k_Nd_N)}.
\end{equation}
\end{linenomath*}
This procedure can be performed further towards shallower layers, and the effective impedance of layer $n$ combined with all its deeper layers would be:
\begin{linenomath*}
\begin{equation}\label{eq:waves_recursive}
	Z_n^\mathrm{eff}=Z_n\frac{Z_{n+1}^\mathrm{eff}\cos(k_nd_n)+iZ_n\sin(k_nd_n)}{iZ_{n+1}^\mathrm{eff}\sin(k_nd_n)+Z_n\cos(k_nd_n)}.
\end{equation}
\end{linenomath*}
By using this recursive relation, we would eventually obtain the effective impedance $Z_1^\mathrm{eff}=\sigma_{1}(x_0)/v_{1}(x_0)$ at the interface 0. This leads to $\sigma_{0}(x_0)/v_{0}(x_0)=Z_1^\mathrm{eff}$ as a result of the continuity of stress and velocity. Substituting this relation into Eq. \ref{eq:waves_right} for the bounding layer 0, we would obtain the total reflection coefficient as:
\begin{linenomath*}
\begin{equation}\label{eq:waves_reflection}
	R=\frac{a_0^-}{a_0^+}=\frac{Z_0-Z_1^\mathrm{eff}}{Z_0+Z_1^\mathrm{eff}},
\end{equation}
\end{linenomath*}
where $Z_0$ is the impedance of the bounding layer 0.

Therefore, upon applying an incident wave $u_0^+(x)=a_0^+e^{-ik_0x}$ on the multilayered medium, the reflected wave can be obtained as $u_0^-(x)=a_0^-e^{ik_0x}$ with a complex amplitude $a_0^-=Ra_0^+$. The reflection coefficient $R$ is given by Eq. \ref{eq:waves_reflection} and the effective impedance therein is calculated using the recursive Eq. \ref{eq:waves_recursive}. We emphasise that the resulting solution takes into account all possible reflections and reverberations since the physical continuity of stress and velocity is considered at every single interface.

\section{Identifying the number of electrode layers using contact setup}\label{appdx:layer_number}

\begin{figure}[ht]
    \centering
    \includegraphics[width=0.9\linewidth]{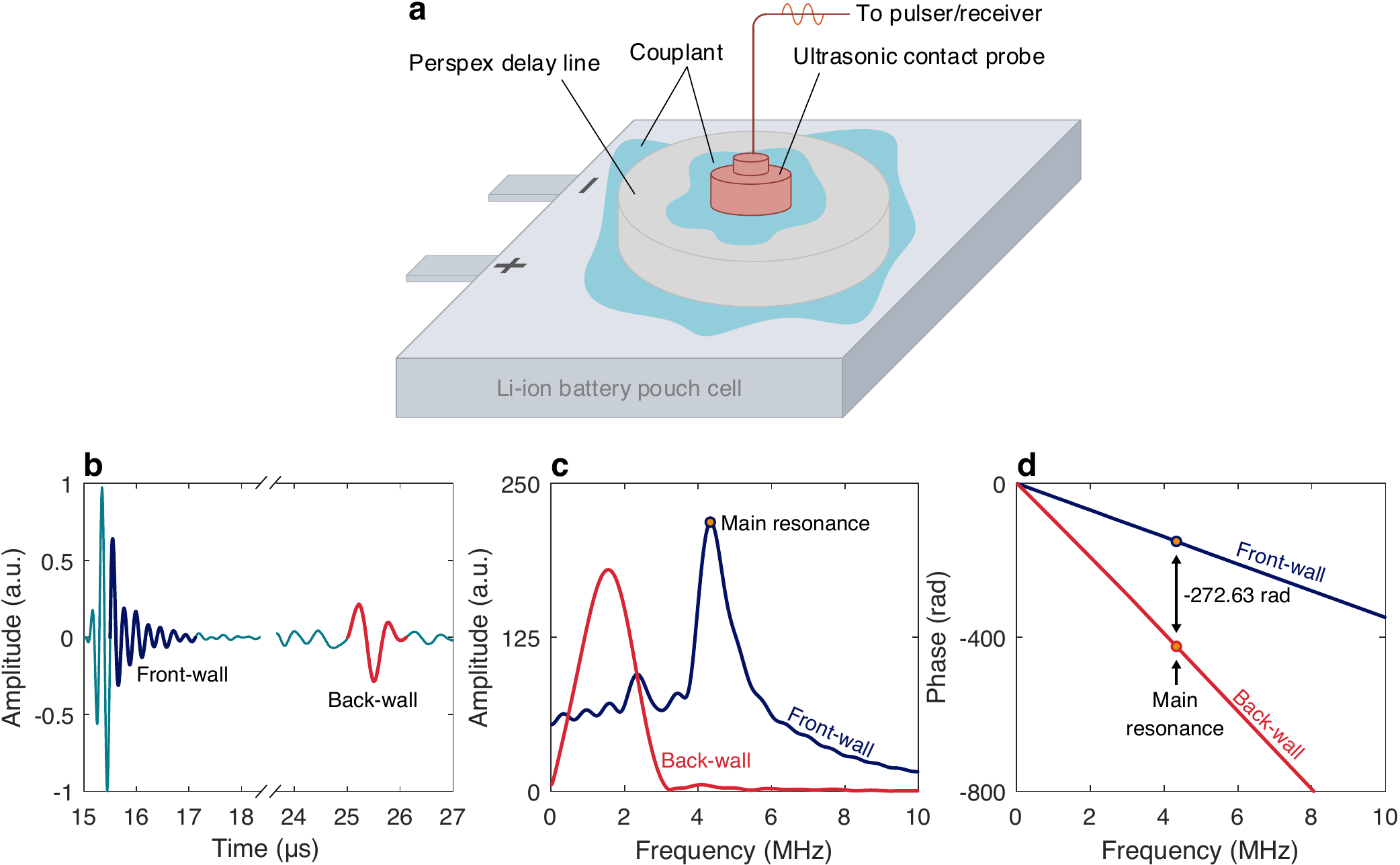}
    \caption{Front- and back-wall reflections from the Kokam 7.5 Ah cell using contact setup. \textbf{a} illustrates the contact setup with a 20-mm thick perspex disc placed between the ultrasonic probe and the cell to capture the front-wall echo of the cell. \textbf{b} shows the time-domain signals acquired with a 5MHz contact probe (V109-RM, Olympus), with the back-wall echo significantly amplified. \textbf{c} and \textbf{d} present the respective amplitude and unwrapped phase spectra of the front- and back-wall echoes.}
    \label{fig:layer_number_contact}
\end{figure}

In section \ref{sec:layer_number}, estimating the number of electrode layers was based on the ultrasonic resonance signals acquired with the immersion setup. Here we perform the same estimation for the Kokam 7.5 Ah cell using the contact setup shown in Fig. \ref{fig:layer_number_contact}a. In the setup, a 20-mm thick perspex disc was used between the probe and the cell in order for the front-wall echo of the cell to be delayed and fully captured by the probe. The front- and back-wall reflections acquired with a 5 MHz probe (V109-RM, Olympus) are provided in Fig. \ref{fig:layer_number_contact}b, with the respective amplitude and phase spectra in c and d. At the main resonant frequency, the phase difference between the two echos is $\phi=-272.63$ rad. After excluding the phase delays caused by the two surface packaging layers of the cell and the phase reversal at the front-wall, the phase change of the wave travelling a round trip through the whole cell is $\phi_\mathrm{total}=-273.88$ rad. The phase change in a resonant element of the cell has been calculated in the main text, which is $\phi_\mathrm{elem}=-5.70$ rad. So, the number $N$ of resonant elements (thus the number of electrode layer pairs) of the cell is given by $N=\phi_\mathrm{total}/\phi_\mathrm{elem}=48.05$. This number is practically the same as that (47.51) obtained in the main text and is very close to the actual number of 48 \cite{Ecker2015,Hales2019}.

\bibliography{references}

\begin{thebibliography}{58}
\providecommand{\natexlab}[1]{#1}
\providecommand{\url}[1]{\texttt{#1}}
\expandafter\ifx\csname urlstyle\endcsname\relax
  \providecommand{\doi}[1]{doi: #1}\else
  \providecommand{\doi}{doi: \begingroup \urlstyle{rm}\Url}\fi

\bibitem[Yoshio et~al.(2009)Yoshio, Brodd, and Kozawa]{yoshio2009lithium}
M.~Yoshio, R.~J. Brodd, and A.~Kozawa.
\newblock \emph{Lithium-ion batteries}, volume~1.
\newblock Springer, 2009.

\bibitem[GOV.UK(2020)]{govuk_2020}
GOV.UK.
\newblock
  \href{https://www.gov.uk/government/news/pm-outlines-his-ten-point-plan-for-a-green-industrial-revolution-for-250000-jobs}{PM
  outlines his ten point plan for a GREEN industrial revolution for 250,000
  jobs}, Nov 2020.
\newblock (Last accessed: 10 Feb 2022).

\bibitem[Reuters(2021)]{eu_ice}
Reuters.
\newblock
  \href{https://www.reuters.com./business/retail-consumer/eu-proposes-effective-ban-new-fossil-fuel-car-sales-2035-2021-07-14/}{EU
  proposes effective ban for new fossil-fuel cars from 2035}, July 2021.
\newblock (Last accessed: 10 Feb 2022).

\bibitem[Nitta et~al.(2015{\natexlab{a}})Nitta, Wu, Lee, and
  Yushin]{nitta2015li}
N.~Nitta, F.~Wu, J.~T. Lee, and G.~Yushin.
\newblock Li-ion battery materials: present and future.
\newblock \emph{Materials today}, 18\penalty0 (5):\penalty0 252--264,
  2015{\natexlab{a}}.

\bibitem[Samsung(2019)]{samsungnote7}
Samsung.
\newblock \href{https://pages.samsung.com/us/note7/recall/index.jsp}{Galaxy
  Note7 safety recall and exchange program}, October 2019.
\newblock (Last accessed: 10 Feb 2022).

\bibitem[Loveridge et~al.(2018)Loveridge, Remy, Kourra, Genieser, Barai, Lain,
  Guo, Amor-Segan, Williams, Amietszajew, Ellis, Bhagat, and
  Greenwood]{batteries4010003}
M.~J. Loveridge, G.~Remy, N.~Kourra, R.~Genieser, A.~Barai, M.~J. Lain, Y.~Guo,
  M.~Amor-Segan, M.~A. Williams, T.~Amietszajew, M.~Ellis, R.~Bhagat, and
  D.~Greenwood.
\newblock Looking deeper into the galaxy (note 7).
\newblock \emph{Batteries}, 4\penalty0 (1), 2018.

\bibitem[Administration(2019)]{tesla2019}
N.~H. T.~S. Administration.
\newblock
  \href{https://static.nhtsa.gov/odi/inv/2019/INIM-DP19005-76719.pdf}{NHTSA
  Defect Petition DP19-005}, October 2019.
\newblock (Last accessed: 10 Feb 2022).

\bibitem[Bloomberg(2020)]{porche_fire}
Bloomberg.
\newblock
  \href{https://www.bloomberg.com/news/articles/2020-02-18/porsche-confirms-taycan-electric-car-caught-fire-in-u-s}{Porsche
  Confirms Taycan Electric Car Caught Fire in U.S.}, February 2020.
\newblock (Last accessed: 10 Feb 2022).

\bibitem[Robinson et~al.(2020)Robinson, Owen, Kok, Maier, Majasan, Braglia,
  Stocker, Amietszajew, Roberts, Bhagat, et~al.]{robinson2020identifying}
J.~B. Robinson, R.~E. Owen, M.~D. Kok, M.~Maier, J.~Majasan, M.~Braglia,
  R.~Stocker, T.~Amietszajew, A.~J. Roberts, R.~Bhagat, et~al.
\newblock Identifying defects in li-ion cells using ultrasound acoustic
  measurements.
\newblock \emph{Journal of The Electrochemical Society}, 167\penalty0
  (12):\penalty0 120530, 2020.

\bibitem[Zhang and Lee(2011)]{zhang2011review}
J.~Zhang and J.~Lee.
\newblock A review on prognostics and health monitoring of li-ion battery.
\newblock \emph{Journal of Power Sources}, 196\penalty0 (15):\penalty0
  6007--6014, 2011.

\bibitem[Chen et~al.(2018)Chen, L{\"u}, Lin, Li, and Pan]{chen2018new}
L.~Chen, Z.~L{\"u}, W.~Lin, J.~Li, and H.~Pan.
\newblock A new state-of-health estimation method for lithium-ion batteries
  through the intrinsic relationship between ohmic internal resistance and
  capacity.
\newblock \emph{Measurement}, 116:\penalty0 586--595, 2018.

\bibitem[Li et~al.(2020)Li, Yuan, Li, and Wang]{li2020state}
X.~Li, C.~Yuan, X.~Li, and Z.~Wang.
\newblock State of health estimation for li-ion battery using incremental
  capacity analysis and gaussian process regression.
\newblock \emph{Energy}, 190:\penalty0 116467, 2020.

\bibitem[Samsung(2021)]{samsung2}
Samsung.
\newblock
  \href{https://www.samsung.com/latin_en/note7-press-conference-detail/}{8-Point
  Battery Safety Check}, November 2021.
\newblock (Last accessed: 10 Feb 2022).

\bibitem[McBreen(2009)]{mcbreen2009application}
J.~McBreen.
\newblock The application of synchrotron techniques to the study of lithium-ion
  batteries.
\newblock \emph{Journal of Solid State Electrochemistry}, 13\penalty0
  (7):\penalty0 1051--1061, 2009.

\bibitem[Finegan et~al.(2015)Finegan, Scheel, Robinson, Tjaden, Hunt, Mason,
  Millichamp, Di~Michiel, Offer, Hinds, et~al.]{finegan2015operando}
D.~P. Finegan, M.~Scheel, J.~B. Robinson, B.~Tjaden, I.~Hunt, T.~J. Mason,
  J.~Millichamp, M.~Di~Michiel, G.~J. Offer, G.~Hinds, et~al.
\newblock In-operando high-speed tomography of lithium-ion batteries during
  thermal runaway.
\newblock \emph{Nature Communications}, 6\penalty0 (1):\penalty0 1--10, 2015.

\bibitem[Vanpeene et~al.(2019)Vanpeene, Villanova, King, Lestriez, Maire, and
  Rou{\'e}]{vanpeene2019dynamics}
V.~Vanpeene, J.~Villanova, A.~King, B.~Lestriez, E.~Maire, and L.~Rou{\'e}.
\newblock Dynamics of the morphological degradation of si-based anodes for
  li-ion batteries characterized by in situ synchrotron x-ray tomography.
\newblock \emph{Advanced Energy Materials}, 9\penalty0 (18):\penalty0 1803947,
  2019.

\bibitem[Ebner et~al.(2013)Ebner, Marone, Stampanoni, and
  Wood]{ebner2013visualization}
M.~Ebner, F.~Marone, M.~Stampanoni, and V.~Wood.
\newblock Visualization and quantification of electrochemical and mechanical
  degradation in li ion batteries.
\newblock \emph{Science}, 342\penalty0 (6159):\penalty0 716--720, 2013.

\bibitem[Frisco et~al.(2016)Frisco, Kumar, Whitacre, and
  Litster]{frisco2016understanding}
S.~Frisco, A.~Kumar, J.~F. Whitacre, and S.~Litster.
\newblock Understanding li-ion battery anode degradation and pore morphological
  changes through nano-resolution x-ray computed tomography.
\newblock \emph{Journal of The Electrochemical Society}, 163\penalty0
  (13):\penalty0 A2636, 2016.

\bibitem[Taiwo et~al.(2017)Taiwo, Finegan, Paz-Garcia, Eastwood, Bodey, Rau,
  Hall, Brett, Lee, and Shearing]{taiwo2017investigating}
O.~O. Taiwo, D.~P. Finegan, J.~Paz-Garcia, D.~S. Eastwood, A.~Bodey, C.~Rau,
  S.~Hall, D.~J. Brett, P.~D. Lee, and P.~R. Shearing.
\newblock Investigating the evolving microstructure of lithium metal electrodes
  in 3d using x-ray computed tomography.
\newblock \emph{Physical Chemistry Chemical Physics}, 19\penalty0
  (33):\penalty0 22111--22120, 2017.

\bibitem[Villevieille et~al.(2010)Villevieille, Boinet, and
  Monconduit]{villevieille2010direct}
C.~Villevieille, M.~Boinet, and L.~Monconduit.
\newblock Direct evidence of morphological changes in conversion type
  electrodes in li-ion battery by acoustic emission.
\newblock \emph{Electrochemistry Communications}, 12\penalty0 (10):\penalty0
  1336--1339, 2010.

\bibitem[Ladpli et~al.(2018)Ladpli, Kopsaftopoulos, and
  Chang]{ladpli2018estimating}
P.~Ladpli, F.~Kopsaftopoulos, and F.-K. Chang.
\newblock Estimating state of charge and health of lithium-ion batteries with
  guided waves using built-in piezoelectric sensors/actuators.
\newblock \emph{Journal of Power Sources}, 384:\penalty0 342--354, 2018.

\bibitem[Hsieh et~al.(2015)Hsieh, Bhadra, Hertzberg, Gjeltema, Goy, Fleischer,
  and Steingart]{hsieh2015electrochemical}
A.~Hsieh, S.~Bhadra, B.~Hertzberg, P.~Gjeltema, A.~Goy, J.~W. Fleischer, and
  D.~A. Steingart.
\newblock Electrochemical-acoustic time of flight: in operando correlation of
  physical dynamics with battery charge and health.
\newblock \emph{Energy \& Environmental Science}, 8\penalty0 (5):\penalty0
  1569--1577, 2015.

\bibitem[Gold et~al.(2017)Gold, Bach, Virsik, Schmitt, M{\"u}ller, Staab, and
  Sextl]{gold2017probing}
L.~Gold, T.~Bach, W.~Virsik, A.~Schmitt, J.~M{\"u}ller, T.~E. Staab, and
  G.~Sextl.
\newblock Probing lithium-ion batteries' state-of-charge using ultrasonic
  transmission--concept and laboratory testing.
\newblock \emph{Journal of Power Sources}, 343:\penalty0 536--544, 2017.

\bibitem[Davies et~al.(2017)Davies, Knehr, Van~Tassell, Hodson, Biswas, Hsieh,
  and Steingart]{davies2017state}
G.~Davies, K.~W. Knehr, B.~Van~Tassell, T.~Hodson, S.~Biswas, A.~G. Hsieh, and
  D.~A. Steingart.
\newblock State of charge and state of health estimation using electrochemical
  acoustic time of flight analysis.
\newblock \emph{Journal of The Electrochemical Society}, 164\penalty0
  (12):\penalty0 A2746, 2017.

\bibitem[Hashin and Shtrikman(1963)]{HASHIN1963127}
Z.~Hashin and S.~Shtrikman.
\newblock A variational approach to the theory of the elastic behaviour of
  multiphase materials.
\newblock \emph{Journal of the Mechanics and Physics of Solids}, 11\penalty0
  (2):\penalty0 127--140, 1963.

\bibitem[Bommier et~al.(2020)Bommier, Chang, Lu, Yeung, Davies, Mohr, Williams,
  and Steingart]{bommier2020operando}
C.~Bommier, W.~Chang, Y.~Lu, J.~Yeung, G.~Davies, R.~Mohr, M.~Williams, and
  D.~Steingart.
\newblock In operando acoustic detection of lithium metal plating in commercial
  licoo2/graphite pouch cells.
\newblock \emph{Cell Reports Physical Science}, 1\penalty0 (4):\penalty0
  100035, 2020.

\bibitem[Chang et~al.(2020)Chang, Mohr, Kim, Raj, Davies, Denner, Park, and
  Steingart]{chang2020measuring}
W.~Chang, R.~Mohr, A.~Kim, A.~Raj, G.~Davies, K.~Denner, J.~H. Park, and
  D.~Steingart.
\newblock Measuring effective stiffness of li-ion batteries via acoustic signal
  processing.
\newblock \emph{Journal of Materials Chemistry A}, 8\penalty0 (32):\penalty0
  16624--16635, 2020.

\bibitem[Robinson et~al.(2019{\natexlab{a}})Robinson, Maier, Alster, Compton,
  Brett, and Shearing]{robinson2019spatially}
J.~B. Robinson, M.~Maier, G.~Alster, T.~Compton, D.~J. Brett, and P.~R.
  Shearing.
\newblock Spatially resolved ultrasound diagnostics of li-ion battery
  electrodes.
\newblock \emph{Physical Chemistry Chemical Physics}, 21\penalty0
  (12):\penalty0 6354--6361, 2019{\natexlab{a}}.

\bibitem[Bauermann et~al.(2020)Bauermann, Mesquita, Bischoff, Drews, Fitz,
  Heuer, and Biro]{BAUERMANN}
L.~P. Bauermann, L.~Mesquita, C.~Bischoff, M.~Drews, O.~Fitz, A.~Heuer, and
  D.~Biro.
\newblock Scanning acoustic microscopy as a non-destructive imaging tool to
  localize defects inside battery cells.
\newblock \emph{Journal of Power Sources Advances}, 6:\penalty0 100035, 2020.

\bibitem[Ecker et~al.(2015)Ecker, Tran, Dechent, K{\"{a}}bitz, Warnecke, and
  Sauer]{Ecker2015}
M.~Ecker, T.~K.~D. Tran, P.~Dechent, S.~K{\"{a}}bitz, A.~Warnecke, and D.~U.
  Sauer.
\newblock {Parameterization of a Physico-Chemical Model of a Lithium-Ion
  Battery: I. Determination of Parameters}.
\newblock \emph{Journal of The Electrochemical Society}, 162\penalty0
  (9):\penalty0 A1836--A1848, 2015.

\bibitem[Hales et~al.(2019)Hales, Diaz, Marzook, Zhao, Patel, and
  Offer]{Hales2019}
A.~Hales, L.~B. Diaz, M.~W. Marzook, Y.~Zhao, Y.~Patel, and G.~J. Offer.
\newblock {The Cell Cooling Coefficient: A Standard to Define Heat Rejection
  from Lithium-Ion Batteries}.
\newblock \emph{Journal of The Electrochemical Society}, 166\penalty0
  (12):\penalty0 A2383--A2395, 2019.

\bibitem[Robinson et~al.(2019{\natexlab{b}})Robinson, Pham, Kok, Heenan, Brett,
  and Shearing]{Robinson2019}
J.~B. Robinson, M.~Pham, M.~D. Kok, T.~M. Heenan, D.~J. Brett, and P.~R.
  Shearing.
\newblock {Examining the Cycling Behaviour of Li-Ion Batteries Using Ultrasonic
  Time-of-Flight Measurements}.
\newblock \emph{Journal of Power Sources}, 444\penalty0 (September):\penalty0
  227318, 2019{\natexlab{b}}.

\bibitem[Pham et~al.(2020)Pham, Darst, Finegan, Robinson, Heenan, Kok,
  Iacoviello, Owen, Walker, Magdysyuk, Connolley, Darcy, Hinds, Brett, and
  Shearing]{Pham2020}
M.~T. Pham, J.~J. Darst, D.~P. Finegan, J.~B. Robinson, T.~M. Heenan, M.~D.
  Kok, F.~Iacoviello, R.~E. Owen, W.~Q. Walker, O.~V. Magdysyuk, T.~Connolley,
  E.~Darcy, G.~Hinds, D.~J. Brett, and P.~R. Shearing.
\newblock {Correlative acoustic time-of-flight spectroscopy and X-ray imaging
  to investigate gas-induced delamination in lithium-ion pouch cells during
  thermal runaway}.
\newblock \emph{Journal of Power Sources}, 470\penalty0 (March):\penalty0
  228039, 2020.

\bibitem[Hunt et~al.(2016)Hunt, Zhao, Patel, and Offer]{Hunt2016}
I.~A. Hunt, Y.~Zhao, Y.~Patel, and J.~Offer.
\newblock {Surface Cooling Causes Accelerated Degradation Compared to Tab
  Cooling for Lithium-Ion Pouch Cells}.
\newblock \emph{Journal of The Electrochemical Society}, 163\penalty0
  (9):\penalty0 A1846--A1852, 2016.

\bibitem[Biot(1956{\natexlab{a}})]{biot1}
M.~A. Biot.
\newblock Theory of propagation of elastic waves in a fluid-saturated porous
  solid. i. lower frequency range.
\newblock \emph{The Journal of the Acoustical Society of America}, 28\penalty0
  (2):\penalty0 168--178, 1956{\natexlab{a}}.

\bibitem[Biot(1956{\natexlab{b}})]{biot2}
M.~A. Biot.
\newblock Theory of propagation of elastic waves in a fluid-saturated porous
  solid. ii. higher frequency range.
\newblock \emph{The Journal of the Acoustical Society of America}, 28\penalty0
  (2):\penalty0 179--191, 1956{\natexlab{b}}.

\bibitem[Martinez-Cisneros et~al.(2016)Martinez-Cisneros, Antonelli, Levenfeld,
  Varez, and Sanchez]{Martinez-Cisneros2016}
C.~Martinez-Cisneros, C.~Antonelli, B.~Levenfeld, A.~Varez, and J.~Y. Sanchez.
\newblock {Evaluation of polyolefin-based macroporous separators for high
  temperature Li-ion batteries}.
\newblock \emph{Electrochimica Acta}, 216:\penalty0 68--78, 2016.

\bibitem[Ledbetter(1980)]{Ledbetter1980}
H.~M. Ledbetter.
\newblock {Sound velocities and elastic-constant averaging for polycrystalline
  copper}.
\newblock \emph{Journal of Physics D: Applied Physics}, 13\penalty0
  (10):\penalty0 1879--84, 1980.

\bibitem[Davies(2018)]{DaviesThesis}
G.~Davies.
\newblock \emph{Characterization of batteries using ultrasound: applications
  for battery management and structural determination}.
\newblock PhD thesis, Princeton University, 2018.

\bibitem[Kube and de~Jong(2016)]{Kube2016}
C.~M. Kube and M.~de~Jong.
\newblock {Elastic constants of polycrystals with generally anisotropic
  crystals}.
\newblock \emph{Journal of Applied Physics}, 120\penalty0 (16):\penalty0
  165105, oct 2016.

\bibitem[Qi et~al.(2014)Qi, Hector, James, and Kim]{Qi2014}
Y.~Qi, L.~G. Hector, C.~James, and K.~J. Kim.
\newblock {Lithium Concentration Dependent Elastic Properties of Battery
  Electrode Materials from First Principles Calculations}.
\newblock \emph{Journal of The Electrochemical Society}, 161\penalty0
  (11):\penalty0 F3010--F3018, 2014.

\bibitem[Koyama et~al.(2006)Koyama, Chin, Rhyner, Holman, Hall, and
  Chiang]{Koyama2006}
Y.~Koyama, T.~E. Chin, U.~Rhyner, R.~K. Holman, S.~R. Hall, and Y.~M. Chiang.
\newblock {Harnessing the actuation potential of solid-state intercalation
  compounds}.
\newblock \emph{Advanced Functional Materials}, 16\penalty0 (4):\penalty0
  492--498, 2006.

\bibitem[Plona(1980)]{plona1980observation}
T.~J. Plona.
\newblock Observation of a second bulk compressional wave in a porous medium at
  ultrasonic frequencies.
\newblock \emph{Applied Physics Letters}, 36\penalty0 (4):\penalty0 259--261,
  1980.

\bibitem[Lu et~al.(2020)Lu, Bertei, Finegan, Tan, Daemi, Weaving, O’Regan,
  Heenan, Hinds, Kendrick, et~al.]{lu20203d}
X.~Lu, A.~Bertei, D.~P. Finegan, C.~Tan, S.~R. Daemi, J.~S. Weaving, K.~B.
  O’Regan, T.~M. Heenan, G.~Hinds, E.~Kendrick, et~al.
\newblock 3d microstructure design of lithium-ion battery electrodes assisted
  by x-ray nano-computed tomography and modelling.
\newblock \emph{Nature communications}, 11\penalty0 (1):\penalty0 1--13, 2020.

\bibitem[Ma et~al.(2019)Ma, Fu, Battaglia, and Prasher]{ma2019microrheological}
F.~Ma, Y.~Fu, V.~Battaglia, and R.~Prasher.
\newblock Microrheological modeling of lithium ion battery anode slurry.
\newblock \emph{Journal of Power Sources}, 438:\penalty0 226994, 2019.

\bibitem[Atkinson and Kytömaa(1992)]{ATKINSON1992577}
C.~Atkinson and H.~Kytömaa.
\newblock Acoustic wave speed and attenuation in suspensions.
\newblock \emph{International Journal of Multiphase Flow}, 18\penalty0
  (4):\penalty0 577--592, 1992.

\bibitem[Auld(1973)]{auld1973acoustic}
B.~A. Auld.
\newblock \emph{Acoustic fields and waves in solids}.
\newblock John Wiley \& Sons, 1973.

\bibitem[Brekhovskikh(1980)]{Brekhovskikh1980}
L.~Brekhovskikh.
\newblock \emph{{Waves in layered media}}.
\newblock Academic Press, New York, NY, 1980.

\bibitem[Nitta et~al.(2015{\natexlab{b}})Nitta, Wu, Lee, and Yushin]{Nitta2015}
N.~Nitta, F.~Wu, J.~T. Lee, and G.~Yushin.
\newblock {Li-ion battery materials: Present and future}.
\newblock \emph{Materials Today}, 18\penalty0 (5):\penalty0 252--264,
  2015{\natexlab{b}}.

\bibitem[Wu et~al.(2019)Wu, Song, Zhang, Hu, Li, Li, Zhang, and Zhang]{Wu2019}
X.~Wu, K.~Song, X.~Zhang, N.~Hu, L.~Li, W.~Li, L.~Zhang, and H.~Zhang.
\newblock {Safety issues in lithium ion batteries: Materials and cell design}.
\newblock \emph{Frontiers in Energy Research}, 7\penalty0 (JUL):\penalty0
  1--17, 2019.

\bibitem[Smith et~al.(2018)Smith, Nelson, Mienczakowski, and Wilcox]{Smith2018}
R.~A. Smith, L.~J. Nelson, M.~J. Mienczakowski, and P.~D. Wilcox.
\newblock {Ultrasonic Analytic-Signal Responses from Polymer-Matrix Composite
  Laminates}.
\newblock \emph{IEEE Transactions on Ultrasonics, Ferroelectrics, and Frequency
  Control}, 65\penalty0 (2):\penalty0 231--243, 2018.

\bibitem[Li et~al.(2021)Li, Kirkaldy, Zhang, Gopalakrishnan, Amietszajew, Diaz,
  Barreras, Shams, Hua, Patel, Offer, and Marinescu]{LI2021229594}
S.~Li, N.~Kirkaldy, C.~Zhang, K.~Gopalakrishnan, T.~Amietszajew, L.~B. Diaz,
  J.~V. Barreras, M.~Shams, X.~Hua, Y.~Patel, G.~J. Offer, and M.~Marinescu.
\newblock Optimal cell tab design and cooling strategy for cylindrical
  lithium-ion batteries.
\newblock \emph{Journal of Power Sources}, 492:\penalty0 229594, 2021.

\bibitem[Zhao et~al.(2018)Zhao, Patel, Zhang, and Offer]{zhao2018modeling}
Y.~Zhao, Y.~Patel, T.~Zhang, and G.~J. Offer.
\newblock Modeling the effects of thermal gradients induced by tab and surface
  cooling on lithium ion cell performance.
\newblock \emph{Journal of The Electrochemical Society}, 165\penalty0
  (13):\penalty0 A3169, 2018.

\bibitem[Gor et~al.(2014)Gor, Cannarella, Pr{\'{e}}vost, and Arnold]{Gor2014}
G.~Y. Gor, J.~Cannarella, J.~H. Pr{\'{e}}vost, and C.~B. Arnold.
\newblock {A Model for the Behavior of Battery Separators in Compression at
  Different Strain/Charge Rates}.
\newblock \emph{Journal of The Electrochemical Society}, 161\penalty0
  (11):\penalty0 F3065--F3071, 2014.

\bibitem[Biot and Willis(1957)]{Biot1957}
M.~A. Biot and D.~G. Willis.
\newblock {The Elastic Coefficients of the Theory of Consolidation}.
\newblock \emph{Journal of Applied Mechanics}, 24\penalty0 (4):\penalty0
  594--601, 1957.

\bibitem[Jocker and Smeulders(2009)]{Jocker2009}
J.~Jocker and D.~Smeulders.
\newblock {Ultrasonic measurements on poroelastic slabs: Determination of
  reflection and transmission coefficients and processing for Biot input
  parameters}.
\newblock \emph{Ultrasonics}, 49\penalty0 (3):\penalty0 319--330, 2009.

\bibitem[Zhao et~al.(1989)Zhao, Tandon, and Weng]{Zhao1989}
Y.~H. Zhao, G.~P. Tandon, and G.~J. Weng.
\newblock {Elastic moduli for a class of porous materials}.
\newblock \emph{Acta Mechanica}, 76\penalty0 (1):\penalty0 105--131, 1989.

\bibitem[B{\'{e}}doui et~al.(2004)B{\'{e}}doui, Diani, and
  R{\'{e}}gnier]{Bedoui2004}
F.~B{\'{e}}doui, J.~Diani, and G.~R{\'{e}}gnier.
\newblock {Micromechanical modeling of elastic properties in polyolefins}.
\newblock \emph{Polymer}, 45\penalty0 (7):\penalty0 2433--2442, 2004.

\end{thebibliography}

\end{document}